\RequirePackage{fix-cm}
\documentclass{svjour3}                     
\smartqed  
\usepackage{graphicx}
\usepackage{amsmath}
\usepackage{amssymb}
\usepackage{bm}

\newcommand{\pan}{{\tt Pandurata}}

\begin{document}

\title{The Collisional Penrose Process}


\author{Jeremy D.\ Schnittman}

\authorrunning{Schnittman} 

\institute{NASA GSFC \\
  8800 Greenbelt Road, mail code 663 \\
  Greenbelt, MD 20771 \\
              \email{jeremy.schnittman@nasa.gov}
}

\date{Received: date / Accepted: date}

\maketitle

\begin{abstract}
Shortly after the discovery of the Kerr metric in 1963, it was
realized that a region existed outside of the black hole's event
horizon where no time-like observer could remain stationary. In 1969,
Roger Penrose showed that particles within this ergosphere region
could possess negative energy, as measured by an observer at
infinity. When captured by the horizon, these negative energy
particles essentially extract mass and angular momentum from the black
hole. While the decay of a single particle within the ergosphere is
not a particularly efficient means of energy extraction, the
{\it collision} of multiple particles can reach arbitrarily high
center-of-mass energy in the limit of extremal black hole spin. The
resulting particles can escape with high efficiency, potentially
serving as a probe of high-energy particle physics as well as general
relativity. In this paper, we briefly review the history of the field
and highlight a specific astrophysical application of the
collisional Penrose process: the potential to enhance annihilation of
dark matter particles in the vicinity of a supermassive black hole.

\keywords{black holes \and ergosphere \and Kerr metric}
\end{abstract}

\section{Introduction}\label{sec:intro}
We begin with a brief overview of the Kerr metric for spinning,
stationary black holes \cite{Kerr:1963}. By far the most convenient,
and thus most common form of the Kerr metric is the form derived by
Boyer and Lindquist \cite{Boyer:1967}. In standard spherical
coordinates $(t,r,\theta,\phi)$ the metric is given by 
\begin{equation}
g_{\mu \nu} = \begin{pmatrix}
-\alpha^2+\omega^2 \varpi^2 & 0 & 0 & -\omega\varpi^2 \\
0 & \rho^2/\Delta & 0 & 0 \\
0 & 0 & \rho^2 & 0 \\
-\omega\varpi^2 & 0 & 0 & \varpi^2 \end{pmatrix}  \, .
\end{equation}

This allows for a relatively simple form for the inverse metric:
\begin{equation}
g^{\mu \nu} = \begin{pmatrix}
-1/\alpha^2 & 0 & 0 & -\omega/\alpha^2 \\
0 & \Delta/\rho^2 & 0 & 0 \\
0 & 0 & 1/\rho^2 & 0 \\
-\omega/\alpha^2 & 0 & 0 & 1/\varpi^2-\omega^2/\alpha^2 \end{pmatrix}  \, .
\end{equation}

In geometrized units with $G=c=1$, we have defined the following terms
\begin{subequations}
\begin{eqnarray}\label{eqn:BL_equations_a}
\rho^2 & \equiv & r^2+a^2\cos^2\theta \\ 
\Delta & \equiv & r^2-2Mr+a^2 \\
\alpha^2 & \equiv & \frac{\rho^2\Delta}{\rho^2\Delta+2Mr(a^2+r^2)} \\
\omega & \equiv & \frac{2Mra}{\rho^2\Delta+2Mr(a^2+r^2)} \\
\varpi^2 & \equiv &
\left[\frac{\rho^2\Delta+2Mr(a^2+r^2)}{\rho^2}\right]\sin^2\theta \label{eqn:BL_equations_e}\, .
\end{eqnarray}
\end{subequations}
It is also convenient to define a dimensionless spin parameter $a_\ast \equiv
a/M$ with $0\le a_\ast \le 1$. Black holes with $a_\ast=1$ are called
{\it maximally spinning} or {\it extremal}.

In these coordinates, the event horizon is located on the surface
defined by $\Delta(r)=0$, giving both an inner and outer horizon:
\begin{equation}
r_{\pm} = M \pm M\sqrt{1-a_\ast^2}\, .
\end{equation}
The outer horizon $r_+$ is generally referred to as the event horizon,
while $r_-$ is also known as the Cauchy horizon. 

Consider a coordinate-stationary observer with 4-velocity
$u^\mu=(u^t,0,0,0)$. For the observer to have a time-like (i.e.
physical) trajectory, we require $g_{tt}u^tu^t < 0$, or alternatively:
\begin{equation}
g_{tt} = -\left(1-\frac{2Mr}{r^2+a^2\cos^2\theta}\right) < 0\, ,
\end{equation}
or 
\begin{equation}
r > M+M\sqrt{1-a_\ast^2\cos^2\theta}\, .
\end{equation}

This implies that there is a region outside of $r_{+}$ where no
stationary observer can exist. This space is called the {\it
  ergosphere}, and is bounded by the surface defined by $r_{\rm E} =
M+M\sqrt{1-a_\ast^2\cos^2\theta}$. While the volume of the ergosphere
is greatest for highly-spinning black holes, it exists for any
black hole with $r_E > r_+$, which is satisfied whenever $a_\ast >
0$. This is expected to be the case for all astrophysical black
holes. 

In 1969, Roger Penrose realized that reactions taking
place within the ergosphere could result in particles with negative
energy \cite{Penrose:1969}. What does it mean for a particle to have negative energy?
Locally, any observer with 4-velocity $u^\mu$ measures the energy of a
particle with momentum $p_\mu$ with a simple inner product: $E_{\rm
  obs} = -p_\mu u^\mu$. An observer at rest at infinity has $u^\mu =
[1,0,0,0]$ so we can define the ``energy at infinity'' of a particle
to be $E_\infty\equiv-p_t$. For stationary spacetimes, $p_t$ is an
integral of motion, so a particle's energy at infinity is
conserved. Thus it is possible for a particle to have positive energy,
as measured by an observer in the ergosphere, yet still have $p_t
>0$. Such a particle simply would not be able to escape the
ergosphere, and would rapidly be captured by the black hole (all the
while maintaining Christodoulou's limits \cite{Christodoulou:1970} on
the black hole mass and spin). 

Shortly after Penrose's 1969 paper, some authors proposed that this
remarkable feature of spinning black holes might be responsible for
the high-energy radiation seen from some active galaxies. But a
careful analysis by Bardeen et al. \cite{Bardeen:1972} and Wald
\cite{Wald:1974} showed that it was impossible to attain relativistic
energies due to the Penrose process alone. A particle would have to
emit a daughter particle with energy a significant fraction of the
original particle's rest mass in order for the surviving particle to
itself reach relativistic velocities. And in this case, while the
escaping particle may in fact be moving near the speed of light, it
can only do so by sacrificing its own rest mass energy. 

Quantitatively, Wald \cite{Wald:1974} showed that a particle with
initial energy $E$ and mass $m$ can decay into a particle with energy
$E'$ and $m'$ in the range
\begin{equation}
  \gamma \frac{E}{m} - \gamma v
  \left(\frac{E^2}{m^2}+1\right)^{1/2} \le \frac{E'}{m'} \le 
  \gamma \frac{E}{m} + \gamma v \left(\frac{E^2}{m^2}+1\right)^{1/2}\, .
\end{equation}
Here, $\gamma$ is the Lorentz factor of the daughter particle in the
original particle's rest frame. For a particle falling in from rest at
infinity ($E/m=1$) and breaking into two equal-mass particles with
mass $m'$ in the ergosphere, we see that the lower energy limit can be
negative when $v>2^{-1/2}$. This is equivalent to two daughter
particles each with rest mass of only $35\%$ of the original
particle. In the limit of 2-photon decay, the upper limit is given by
$E'\le E\frac{1+\sqrt{2}}{2}$, corresponding to an efficiency of
$\sim 121\%$. 

Here we should note that our definition of efficiency differs slightly
from some previous works. For example, Piran \& Shaham
\cite{Piran:1977} define the efficiency of the Penrose process by
\begin{equation}
\eta_{\rm P-S} \equiv \frac{E_{\rm esc}-E_{\rm in}}{E_{\rm in}} = \frac{-E_{\rm
    cap}}{E_{\rm in}}\, ,
\end{equation}
where the incoming particle has energy $E_{\rm in}$, the escaping
particle $E_{\rm esc}$, and the particle captured by the black hole
has $E_{\rm cap}$, which is taken to be negative for the Penrose
process. In an attempt to distinguish between the exotic nature of the
Penrose process and more conventional physical processes, we use a
definition of efficiency that compares total energy out with total
energy in:
\begin{equation}
\eta \equiv \frac{E_{\rm esc}}{E_{\rm in}}\, ,
\end{equation}
which will have values exceeding $100\%$ for the Penrose process. 

Now, $121\%$ efficiency in converting mass to energy is nothing to
dismiss lightly. It far exceeds the efficiency of nuclear fusion, and
even exceeds the radiative efficiency of quasars (assuming
Novikov-Thorne thin accretion disks \cite{Novikov:1973} and maximal
spin, efficiency of $40\%$
might be attained \cite{Thorne:1974}). It is the ultimate ``free
lunch:'' getting more energy out than you put in.
Yet it still cannot explain the GeV or even TeV emission seen in
blazars, or the large-scale, relativistic jets seen emerging from
radio galaxies with bulk Lorentz factors 
of 10 or more, and this was one of the outstanding mysteries that
Penrose, Wald, and others at the time were trying to solve
\cite{Penrose:1969,Wald:1974}.  

\begin{figure}
\begin{center}
  \includegraphics[width=0.65\textwidth]{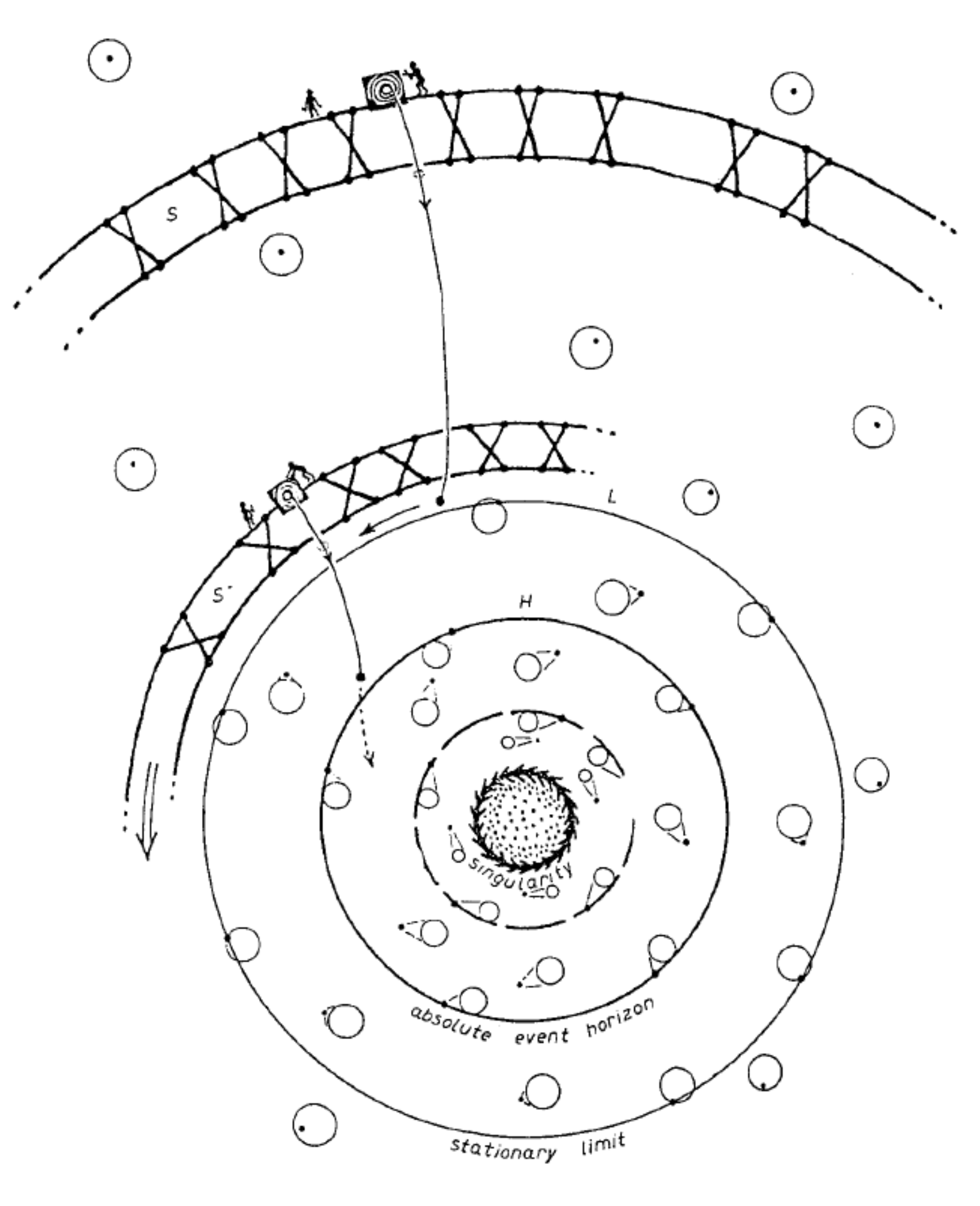}
\caption{\label{fig:powerplant} Potential application of the Penrose process for energy
  extraction from a spinning black hole with macroscopic
  particles. The image depicts concentric rings of superstructures
  orbiting the black hole, with people lowering masses on pulleys to
  extract gravitational energy from the mass. Inside the ergosphere
  (marked ``static limit'' in this figure), these packages can attain
  negative energy, and thus greater than unity efficiency in energy
  generation. Also shown in this figure is a sample of causal light
  cones, depicted as circles around dots in the outer region, and
  inward-tilting cones inside the horizon. Reproduced from
  \cite{Penrose:1969}.}
\end{center}
\end{figure}

While not an obvious solution for the origin of relativistic jets, the
Penrose process was seen, perhaps whimsically, as a potential and
exotic source of energy for an extremely advanced civilization living
around a Kerr black hole. This is shown in Figure
\ref{fig:powerplant}, reproduced from Penrose's original 1969 paper. 

Shortly after Wald published his limit of $121\%$, Piran and
collaborators discovered a way to attain an even higher efficiency
from the Penrose process \cite{Piran:1975,Piran:1977}. The crucial
point was to use more than one particle, in order to achieve a much
higher center-of-mass energy in the ergosphere. This approach is known
as the {\it Collisional Penrose Process}. 

The collisional Penrose process is a great deal richer than the simple
decay problem considered by Wald, where it was clear from inspection
what was the geometry required for maximal efficiency. For {\it two}
incoming particles, the range of possible values for the total energy
and momentum are vastly greater. Further expanding the population of
colliding particles to
those with $E_0/m \ne 1$ and non-planar orbits makes the problem
virtually intractable for analytic methods. However, from symmetry
arguments we can limit our focus on specialized planar orbits
around extremal black holes when looking for the highest efficiency
reactions. 

As discussed at length in \cite{Piran:1977}, it is not enough for the
daughter particles to have large values of energy-at-infinity, but
they must also be able to escape the potential of the black hole. 
Following our approach in \cite{Schnittman:2014}, we can write down an
effective potential for geodesic trajectories in the equatorial plan
around a Kerr black hole. From the normalization constraint
$g_{\mu \nu}p^\mu p^\nu=-m^2$ we have
\begin{equation}\label{eqn:Veff}
V_{\rm eff}(r) = k\frac{M}{r}+\frac{\ell^2}{2r^2}
+\frac{1}{2}(-k-\varepsilon^2)\left(1+\frac{a^2}{r^2}\right)
-\frac{M}{r^3}(\ell-a\varepsilon)^2\, ,
\end{equation}
where $\ell$ and $\varepsilon$ are the particle's specific angular
momentum and energy, and $k=0$ for photons and
$k=-1$ for massive particles. 
For a specific choice of $a$, $\ell$, $\varepsilon$, and $k$,
we can solve for the radial turning points by setting 
$V_{\rm eff}(r)=0$. 

\begin{figure}
\includegraphics[width=0.45\textwidth,clip=true]{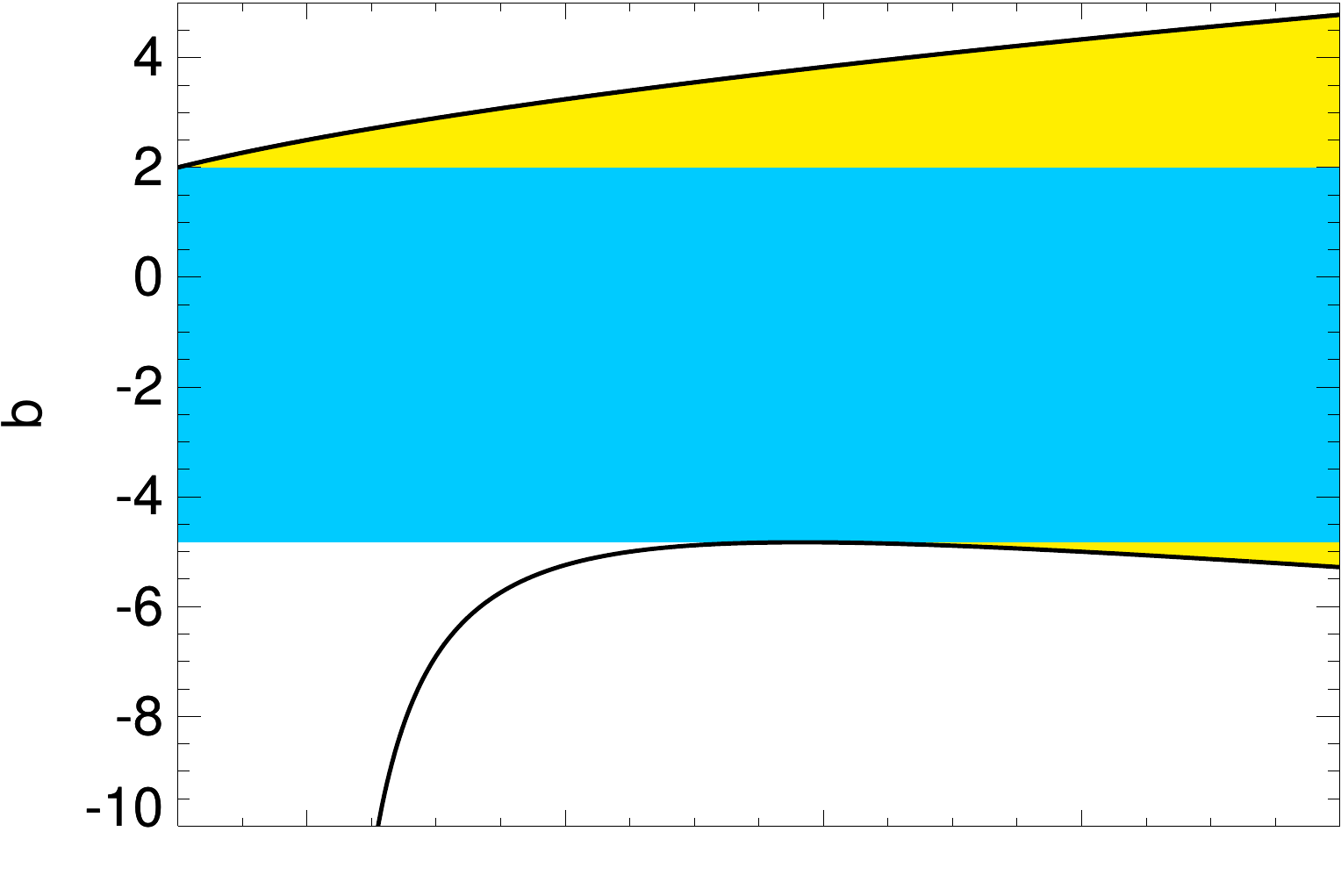}
\includegraphics[width=0.45\textwidth,clip=true]{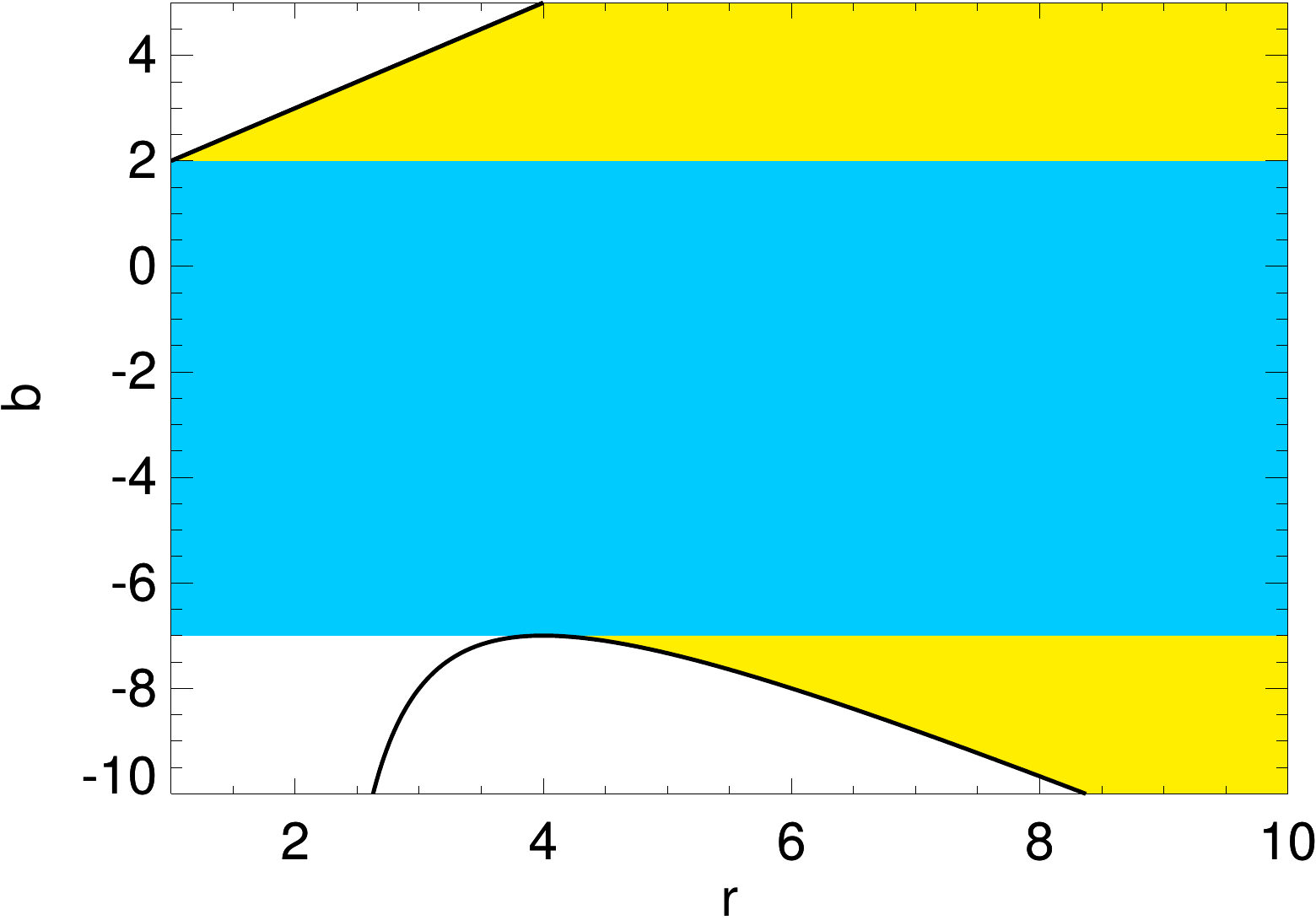}
\caption{\label{fig:Veff}
  Radial turning points in the effective potential $V_{\rm
    eff}(r,b)$ for ({\it left}) massive and ({\it right}) massless
  particles, for a black hole with maximal spin $a_\ast=1$. Any particle in
  the yellow region can escape from the black
  hole, but in the blue regions, only particles with outgoing
  radial velocities can escape. The static limit is located at 
  $r_{\rm E}=2$ and the horizon is at $r_+=1$.}
\end{figure} 

Figure \ref{fig:Veff} shows these turning points as a function of the
impact parameter $b\equiv \ell/\varepsilon$ for both massless and
massive particles, for maximal spin $a_\ast=1$. For the
massive particles we set $\varepsilon=1$, corresponding to a particle
at rest at infinity. One can visualize a massive particle incoming
from the right with $b < -2(1+\sqrt{2})$ or $b > 2$, reflecting
off the centrifugal potential barrier and returning back to infinity (yellow
regions). Alternatively, if the impact parameter is small enough
[i.e., $-2(1+\sqrt{2}) < b < 2$], the particle will get captured by
the black hole. Due to frame-dragging, the cross section for capture
is much greater for incoming particles with negative angular
momentum \cite{Chandra:1992}.

For a given $\ell=p_\phi$ and $\varepsilon=-p_t$, the radial momentum
$p_r$ can be determined from the normalization condition $p_\mu p_\nu
g^{\mu \nu}=k$:
\begin{equation}\label{eqn:p_r}
p_r = \pm [g_{rr}(k-g^{tt}\varepsilon^2+2g^{t\phi}\ell\varepsilon
  -g^{\phi \phi}\ell^2)]^{1/2}\, ,
\end{equation}
and the sign of the root is chosen depending on criterion described
below. The conditions described by equation (\ref{eqn:p_r}) and Figure
\ref{fig:Veff} will be essential in understanding the range of
energies attainable by outgoing particles. Yet before we get to that
calculation in Section \ref{sec:super}, we will give an overview of
the ``BSW'' effect, named after the paper in 2009 by Banados, Silk,
and West \cite{Banados:2009}, which revitalized interest in the field
more than thirty years after the exhaustive analysis of
\cite{Piran:1977}. 

\section{Banados-Silk-West}\label{sec:BSW}
The primary result of BSW is that, for two particles falling in from
rest at infinity, in the limit of extremal spin and collisions close
to the horizon, the center-of-mass energy can reach arbitrarily high
levels. This effect was indeed pointed out in \cite{Piran:1977}, but
those authors recognized it as not particularly interesting from an
astrophysical point of view. Perhaps also because of the numerical
simplicity of BSW, it received a great deal of attention immediately
after publication. In this section, we will first repeat the BSW
calculation, and then summarize a collection of critiques about the
possibility of ever reaching such energies in practice. There have
also been a great number of papers focusing on BSW-type reactions, but
for non-Kerr black holes. In the interest of brevity, we do not
include any discussion of those results in this review, which is
focused specifically on the classical collisional Penrose process. 

\begin{figure}
\begin{center}
  \includegraphics[width=0.45\textwidth]{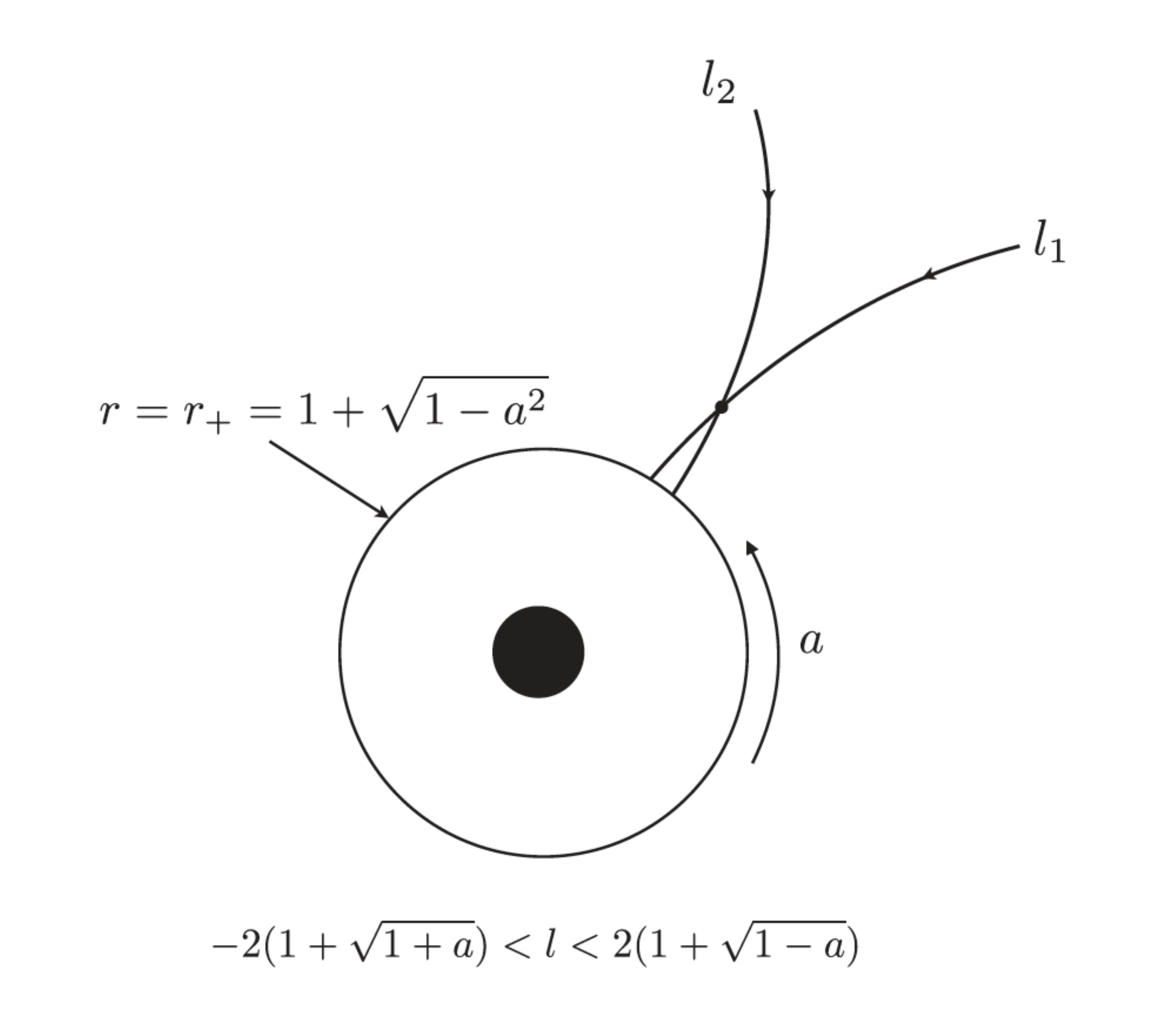}
\caption{\label{fig:BSW_fig1} 
  Schematic picture of two particles falling into a black hole
  with spin $a_\ast$ and colliding near the horizon
  ($r_+$). Reproduced from \cite{Banados:2009}.}
\end{center}
\end{figure}

Figure \ref{fig:BSW_fig1} is reproduced from BSW, showing a schematic
picture of the incoming particles colliding near the horizon. As can
be derived from equation (\ref{eqn:Veff}) and Figure \ref{fig:Veff},
the allowed range for the impact parameter $b$ (same as BSW $l$, as
the energy for particles falling in from infinity is unity) is
\begin{equation}
-2(1+\sqrt{1+a_\ast}) < b < 2(1+\sqrt{1-a})\, .
\end{equation}
We take our two
particles $\mathbf{p}^{(1)}$ and $\mathbf{p}^{(2)}$ to have 4-momentum
components $[-1,p_r^{(1,2)},0,\ell^{(1,2)}]$ with $p_r$ computed as
above in equation (\ref{eqn:p_r}).

\begin{figure}
\begin{center}
  \includegraphics[width=0.55\textwidth]{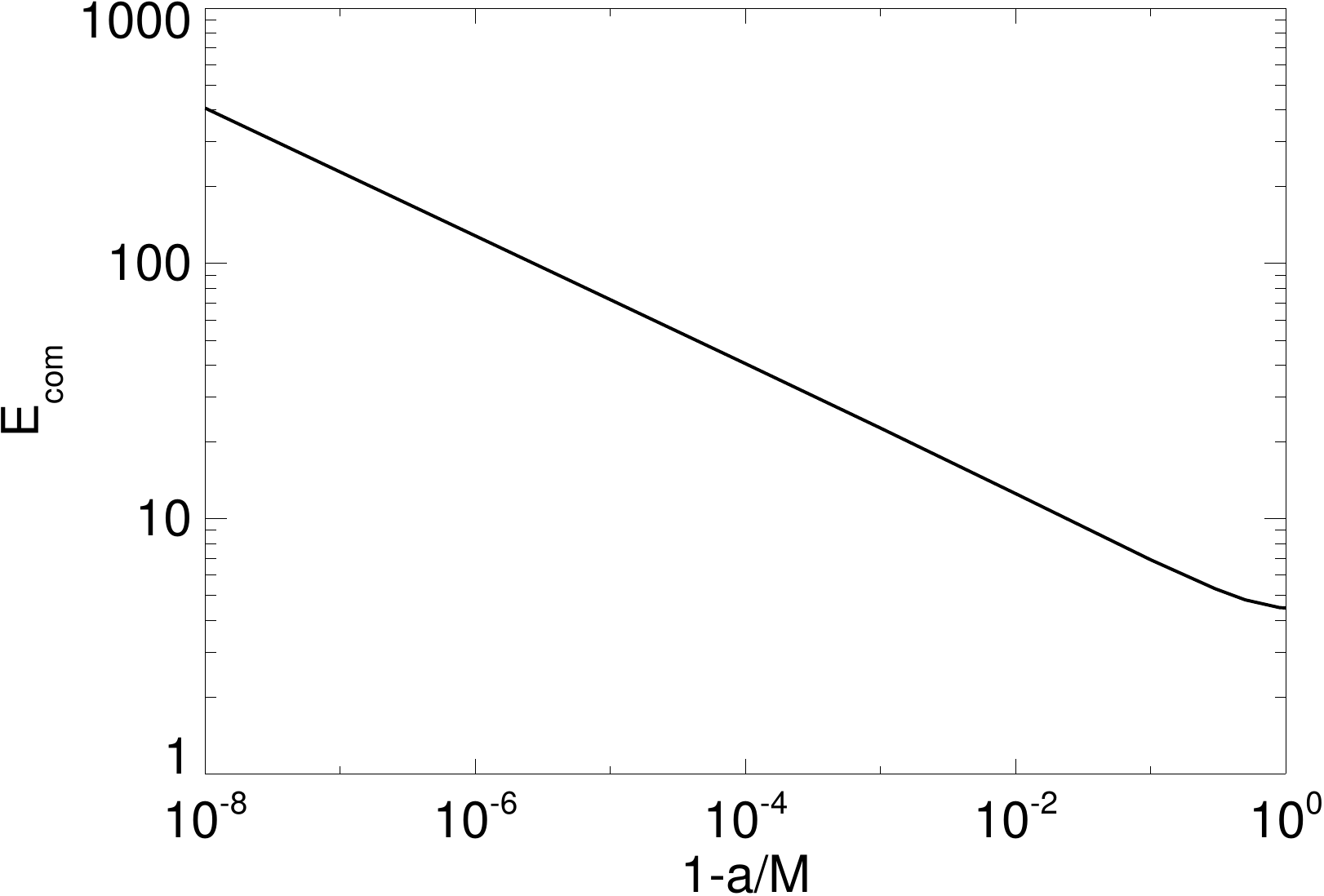}
\caption{\label{fig:Ecom} 
  Center-of-mass energy as a function of spin for two particles
  falling from rest at infinity with critical impact parameters $b=\pm
  2(1+\sqrt{1\mp a})$, colliding near the event horizon.}
\end{center}
\end{figure}

The center-of-mass energy is given by the expression
\begin{equation}\label{eqn:Ecom}
E_{\rm com} = \sqrt{2(1-g^{\mu \nu}p_\mu^{(1)}p_\nu^{(2)})}\, .
\end{equation}
The simplified expression from BSW is 
\begin{eqnarray}\label{eqn:Ecom2}
E_{\rm com}^2 &=& \frac{2m_0^2}{r(r^2-2r+a_\ast^2)} \left[2a^2(1+r) -
  2a(\ell^{(1)}+\ell^{(2)}) -\ell^{(1)}\ell^{(2)}(r-2)+2(r-1)r^2 
  \right. \nonumber\\
& & \left. - \sqrt{2(a-\ell^{(1)})^2-\ell^{(1)2}+2r^2}
\sqrt{2(a-\ell^{(2)})^2-\ell^{(2)2}+2r^2} \right] \, .
\end{eqnarray}
Note that BSW adopts units such that the mass is taken to be unity, and
thus does not appear in equation (\ref{eqn:Ecom2}). The denominator in
equation (\ref{eqn:Ecom2}) is always zero at the horizon, so it may
appear at first glance that the center-of-mass energy always diverges,
regardless of black hole spin. However, from the effective potential
Figure \ref{fig:Veff}, one sees that only a range of allowed values
for $b$ are able to actually reach the horizon. When taking these
limits for $\ell^{(1,2)}$ and taking the location of the collision to
approach the horizon, one finds that the center-of-mass energy is in
fact finite for non-extremal spins. The actual algebraic expression is
rather cumbersome, but in Figure \ref{fig:Ecom} we plot $E_{\rm com}$
for these critical orbits, assuming a collision just outside the event
horizon. Note that while the energy diverges in the limit of $a_\ast
\to 1$, it does so quite slowly, roughly as $E_{\rm com}\sim
(1-a_\ast)^{-1/4}$. 

Equation (\ref{eqn:Ecom2}) simplifies significantly for extremal black
holes with $a_\ast=1$, giving the center-of-mass energy at the horizon as
\cite{Banados:2009}
\begin{equation}
E_{\rm com}^2 = 2m_0^2 \left(\frac{\ell^{(2)}-2}{\ell^{(1)}-2} + 
\frac{\ell^{(2)}-2}{\ell^{(1)}-2}\right)\, .
\end{equation}
Thus if either (but not both) of the particles have the critical
angular momentum of $\ell=2$, the collisional energy diverges. 

This divergence for the extremal case is closely related to the
curvature structure of the horizon. It is well known that the
singularity of the Kerr metric at the horizon is only a coordinate
singularity for Boyer-Lindquist coordinates \cite{Visser:2007}. A whole class of
coordinates exists (e.g., Kerr-Schild, Doran) that do not blow up at
the horizon, and are thus useful for calculating the trajectories of
particles near or across the horizon. However, in the limit of
extremal Kerr, the curvature singularity approaches the horizon at
Boyer-Lindquist $r=M$ (a super-extremal black hole would have the
singularity {\it outside} the horizon, and thus violate the cosmic
censorship conjecture), and thus physical, coordinate-independent
quantities such as the center-of-mass energy can diverge. 

In response to BSW, Lake \cite{Lake:2010} and Gau \& Zhong
\cite{Gao:2011} showed that the c.o.m. energy for
collisions {\it inside} the horizon will generically diverge even for
non-extremal black holes as the
particles approach the inner Cauchy horizon, which is itself outside of the
curvature singularity (they all coincide with $r=M$ in the extremal
limit). Along these lines, ref.\ \cite{Stuchlik:2013} showed that, for
black holes with $a_\ast>1$ (naked singularities), infinite
c.o.m. energy collisions were quite generic. 

While the diverging energy is apparently mathematically possible, in
the wake of BSW, numerous authors raised physical or astrophysical
objections to the proposition of using Kerr black holes to probe
extreme particle energies. Here we provide a brief summary of some of
the more interesting objections. Please see \cite{Harada:2014} for an
excellent review of the topic.

\begin{itemize}
\item Berti et al.\ \cite{Berti:2009} point out two practical
  problems: even in the limit of an initially extremal black hole, a
  single collision would deposit the mass and angular momentum of the
  debris particles, lowering the black hole spin far below the levels
  needed for Planck-scale collisions. Additionally, the critical
  orbits required for diverging $E_{\rm com}$ take an infinite amount
  of proper time to actually reach the horizon. During this time, a
  particle would orbit so many times that it would actually emit a
  significant amount of energy and angular momentum in gravitational
  radiation, in turn reducing the spin of the black hole. For the much
  more astrophysical spin limit of $a_\ast=0.998$ \cite{Thorne:1974},
  the peak energy would be a paltry 6.95 times the rest-mass energy
  \cite{Harada:2011}. 
\item Jacobson \& Sotiriou \cite{Jacobson:2010} show that the scaling
  of $E_{\rm com}$ is extremely weak with the spin. For near-maximal
  $a=1-\epsilon$, they find the peak energy to be $E_{\rm com} \sim
  4.06 \epsilon^{-1/4}$ (see Fig.\ \ref{fig:Ecom} above). Aside from
  this weak scaling restriction, they also show
  that any energy gained by colliding near the horizon will
  necessarily be lost by the redshift of escaping the black hole
  potential. 
\item Harada \& Kimura \cite{Harada:2011} demonstrated similar scaling
  for particles falling in from the inner-most stable circular orbit
  (ISCO). For (non-plunging) particles on ISCO orbits colliding with
  generic particles falling in from infinity, the peak center-of-mass
  energy has an even weaker scaling: $E_{\rm com}\sim 5 \epsilon^{-1/6}$.
\item Bejger {\it et al.} \cite{Bejger:2012} focus on the problem of
  the escaping particle's 
  energy. They agree that an arbitrary center-of-mass energy can be
  achieved, but like \cite{Jacobson:2010}, point out that the reaction products lose much of
  their energy on the way out from the horizon, ultimately limiting
  the efficiency of the process to $129\%$ for equal-mass particles
  falling in from infinity. 
\item Harada {\it et al.} \cite{Harada:2012} carry out a more general
  calculation including non-equal mass particles and Compton
  scattering reactions, yet mistakenly calculate an even smaller upper
  limit of $109\%$ for efficiency in the BSW-type reaction. 
\item Ding {\it et al.} \cite{Ding:2013} are the first to introduce
  the additional limitations that will arise from a quantum theory of
  gravity. By including the effects of a non-commutative spacetime via
  a parameterized effective field theory, they show that the maximum
  center-of-mass energy attainable is of the order of a few thousand
  times the particle rest mass, but depends on the black hole mass (in
  quantum gravity, black holes are no longer scale invariant). 
\item Galajinsky \cite{Galajinsky:2013} repeats the BSW calculation in
  both Boyer-Lindquist and Near-Horizon Extremal Kerr (NHEK)
  coordinates, and surprisingly finds two different answers for the
  maximum c.o.m. energy, with it diverging in the classical
  Boyer-Lindquist approach, but remaining finite in NHEK. This
  apparent paradox is likely due to the order in which various
  diverging limits are taken, and which values are allowed for
  particle trajectories, with the consensus appearing that the
  B-L result is correct \cite{Zaslavskii:2013}. 
\item Patil {\it et al.} \cite{Patil:2015} point out that for
  ultra-high values for the c.o.m. energy, the critical particles must
  come in with such finely-tuned values of angular momentum that they
  take a nearly infinite amount of coordinate time to reach the
  horizon, or even the radius of collision necessary for Planck-scale
  energies. A potential way around this problem with multiple
  collisions was identified by Griv \& Pavlov \cite{Grib:2011}.
\item Like Berti {\it et al.} \cite{Berti:2009}, McCaughey
  \cite{McCaughey:2016} also questions the possibility of an extremal
  black hole existing in nature. In particular, he focuses on the
  problem of Hawking radiation combined with the Penrose process of
  virtual particles in the ergosphere, which have a tendency of
  spinning down the black hole. Unfortunately, that paper does not
  include a quantitative estimate of the physical spin-down rate as a
  function of black hole mass and spin (see below). 
\item Most recently, Hod \cite{Hod:2016} raised yet another problem
  for reaching the highest c.o.m. energies, based on Thorne's classic
  hoop conjecture \cite{Thorne:1972}. Simply put, if you pack enough
  energy into a small enough area, you form a black hole. In the
  context of the BSW process, if this energy is in the form of the
  colliding particles, and they are close enough to the horizon, then
  the two black holes instantly merge, and the daughter particles
  cannot escape, regardless of their nominal energy and angular
  momentum.
\end{itemize}

In the process of compiling this collection of challenges to the
possibility of a divergent c.o.m. energy, two other potential problems
occurred to us, apparently not addressed up to this point in the
literature. The first is the spin-down of the extremal black hole due
to Hawking radiation \cite{Bekenstein:1973,Hawking:1975}. This effect was first explored by Page in 1976
\cite{Page:1976}, and that is still one of the clearest, most
comprehensive analyses of mass and spin evolution due to Hawking
radiation. One of the interesting results from \cite{Page:1976} is
that, for extremal black holes, the evolution is dominated by graviton
losses, orders of magnitude greater than the contribution of photons
or neutrinos. This is perhaps not surprising, as the curvature
singularity is so close to the horizon for $a_\ast=1$. 

Writing $a_\ast = J/M^2$, the expression for spin evolution due to
Hawking radiation is given by 
\begin{equation}\label{eqn:dadt}
\frac{da_\ast}{dt} = -2M^{-3} J \frac{dM}{dt} + M^{-2}\frac{dJ}{dt}
= -2a_\ast \dot{M}/M + \dot{J}/M^2 = M^{-3}a_\ast[2
  f(a_\ast)-g(a_\ast)]\, ,
\end{equation}
where $f(a_\ast)$ and $g(a_\ast)$ are numerical functions tabulated in
\cite{Page:1976}, as a function of the spin parameter and species of
Hawking particle (photons, neutrinos, gravitons). Setting $a_\ast=1$
and $M=10^8 M_\odot$, we find the spindown rate to be 
$\dot{a}_\ast \approx -7\times 10^{-97}$ s$^{-1}$. At that rate, the
spin will remain high enough to allow Planck-scale BSW reactions for a
Hubble time. However, the strong mass scaling means that, for
stellar-mass black holes of $M \sim 10 M_\odot$, the spindown rate
would be on the order of $\dot{a}_\ast \approx 10^{-75}$
s$^{-1}$. While this still sounds extremely small (i.e., the spin will
remain near-extremal for a very long time), let us review the
c.o.m. scaling with spin. 

For spin $a_\ast = 1-\epsilon$, the critical angular momentum (maximum
c.o.m. energy) for incoming particles is $b_{\rm crit} \approx
2(1+\epsilon^{1/2})$. The critical radius for these collisions is at
$r_{\rm crit} \approx 1+2\epsilon^{1/2}$, and the c.o.m. energy scales
like $E_{\rm COM} \approx (r-1)^{-1/2} \approx \epsilon^{-1/4}$
\cite{Berti:2009,Jacobson:2010}. So for an incoming particle with rest
mass on the order of a GeV, in order to reach Planck energies ($\sim
10^{19}$ GeV) the critical spin is $1-a_\ast \lesssim 10^{-76}$. In
other words, an extremal stellar-mass black hole would spin down from
Hawking radiation in under a second (however, see below in
Sec.\ \ref{sec:super} for a less conservative limit on the critical
spin value).

Focusing for now on the supermassive black holes, where Hawking
radiation should not be important, there is however another, more
astrophysical mechanism to spin down the black hole. All astrophysical
black holes are surrounded by a bath of isotropic thermal radiation
from the cosmic microwave background. At a present temperature of 2.73
K, this radiation is far more energetic than the Hawking radiation
from any black hole larger than the mass of the moon ($\sim 10^{-8}
M_\odot$). Furthermore, it is isotropic, so a Kerr black hole will
preferentially absorb photons with negative angular momentum, thereby
accelerating the spin-down process. 

From numerical calculations with the {\tt Pandurata} ray-tracing code
\cite{Schnittman:2013}, we can determine the cross-section of an
extremal black hole to radiation incoming from infinity, and find an
angle-averaged effective radius of 
$r_{\rm eff} \approx \sqrt{23}r_g$, with the gravitational
radius of a black hole defined by $r_g \equiv GM/c^2$. We found
the mean specific angular momentum of a captured photon to be $-1.6
r_g$. So following from equation (\ref{eqn:dadt}) we get
\begin{equation}
\frac{da_\ast}{dt} = -2a_\ast \frac{\dot{M}}{M} + \frac{\dot{J}}{M^2}
= -3.6 \frac{\dot{M}}{M} = 4\times 10^{-36}\left(\frac{T}{2.73\mbox{
    K}}\right)^4 \left(\frac{M}{10^8 M_\odot}\right) \mbox{ s}^{-1}\, ,
\end{equation}
where the final expression comes from the ``accretion'' of the CMB
flux $\dot{M}c^2 = 4\pi r_{\rm eff}^2 \sigma T^4$. A smaller but
similar level of flux is received from the cosmic neutrino
background. 

Clearly, the effect of CMB accretion dominates over Hawking radiation
for any astrophysical black hole. Even for the smallest known black
holes, an initially extremal black hole would spin down well below the
critical BSW/Planck spin value in a tiny fraction of a second. 

In addition to these many critiques and commentaries on BSW, there
has been an even larger number of follow-on papers exploring analogous
effects in non-Kerr black holes. These papers were both within the
limits of general relativity (e.g., Kerr-Newman metric), as well as
alternative theories of 
gravity. However, since this review (and the entire Topical
Collection) is specifically concerned with the Kerr metric, we
consider these alternative approaches to be outside the scope of our
present discussion. 

\section{Super-Penrose Process}\label{sec:super}

One of the most interesting aspects of the post-BSW literature is the
question of the range of energies and escape fraction of the reaction
products. As discussed above, the original limit of Wald
\cite{Wald:1974} was only $121\%$ for the spontaneous decay of a
massive particle into two photons. To better understand the range of
attainable energies and their relative likelihoods, we introduce a
novel graphical representation of the reaction products. In the
interest of tractability of a many-dimensional problem, for this
entire section, unless otherwise stated, we will restrict our analysis
to planar equatorial trajectories for extremal Kerr black holes. 

\begin{figure}
\begin{center}
  \includegraphics[width=0.07\textwidth]{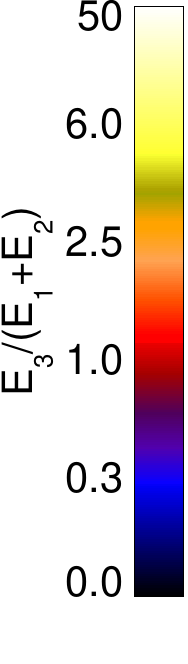}\hspace{2mm}
  \includegraphics[width=0.28\textwidth]{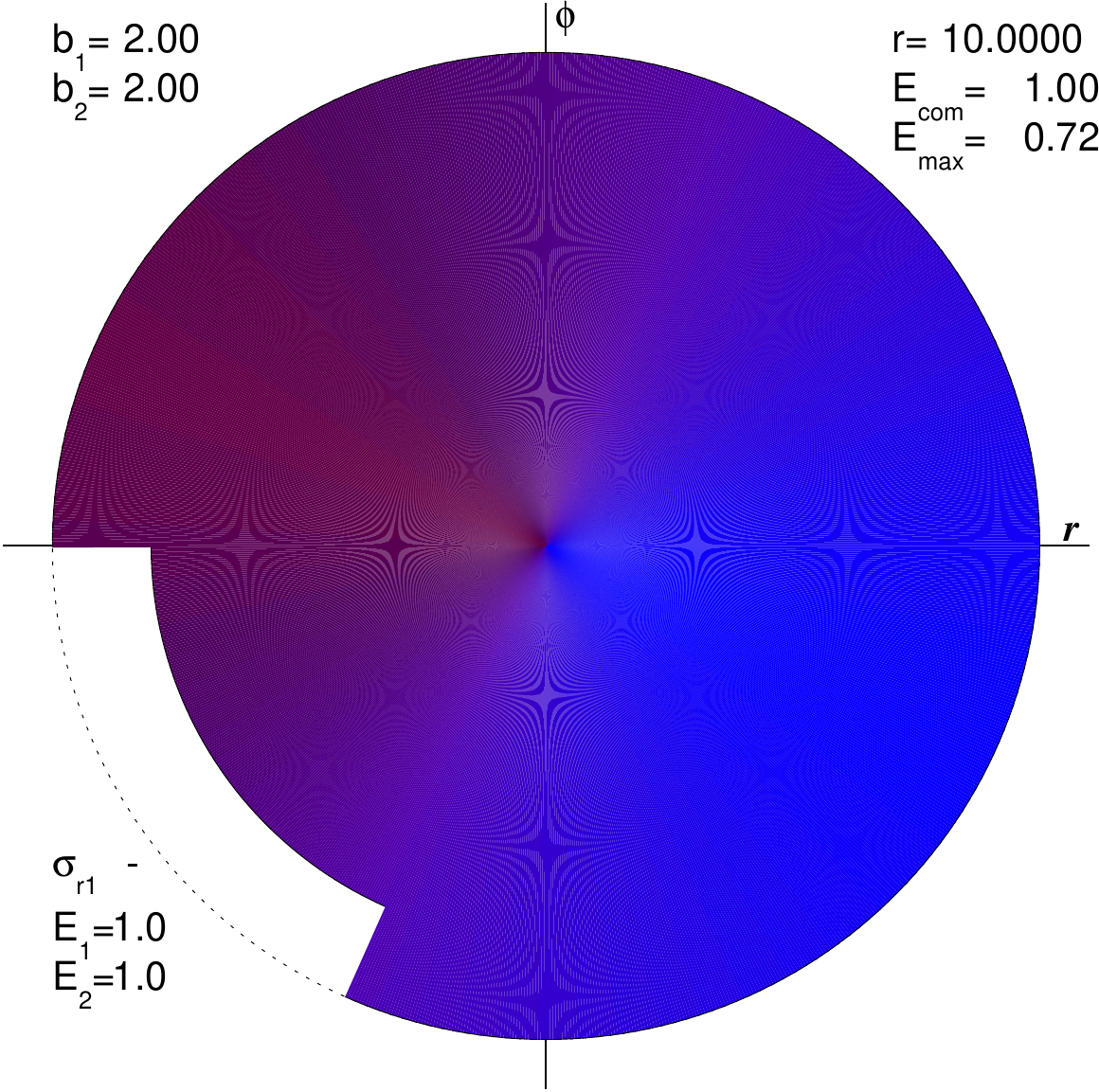}\hspace{2mm}
  \includegraphics[width=0.28\textwidth]{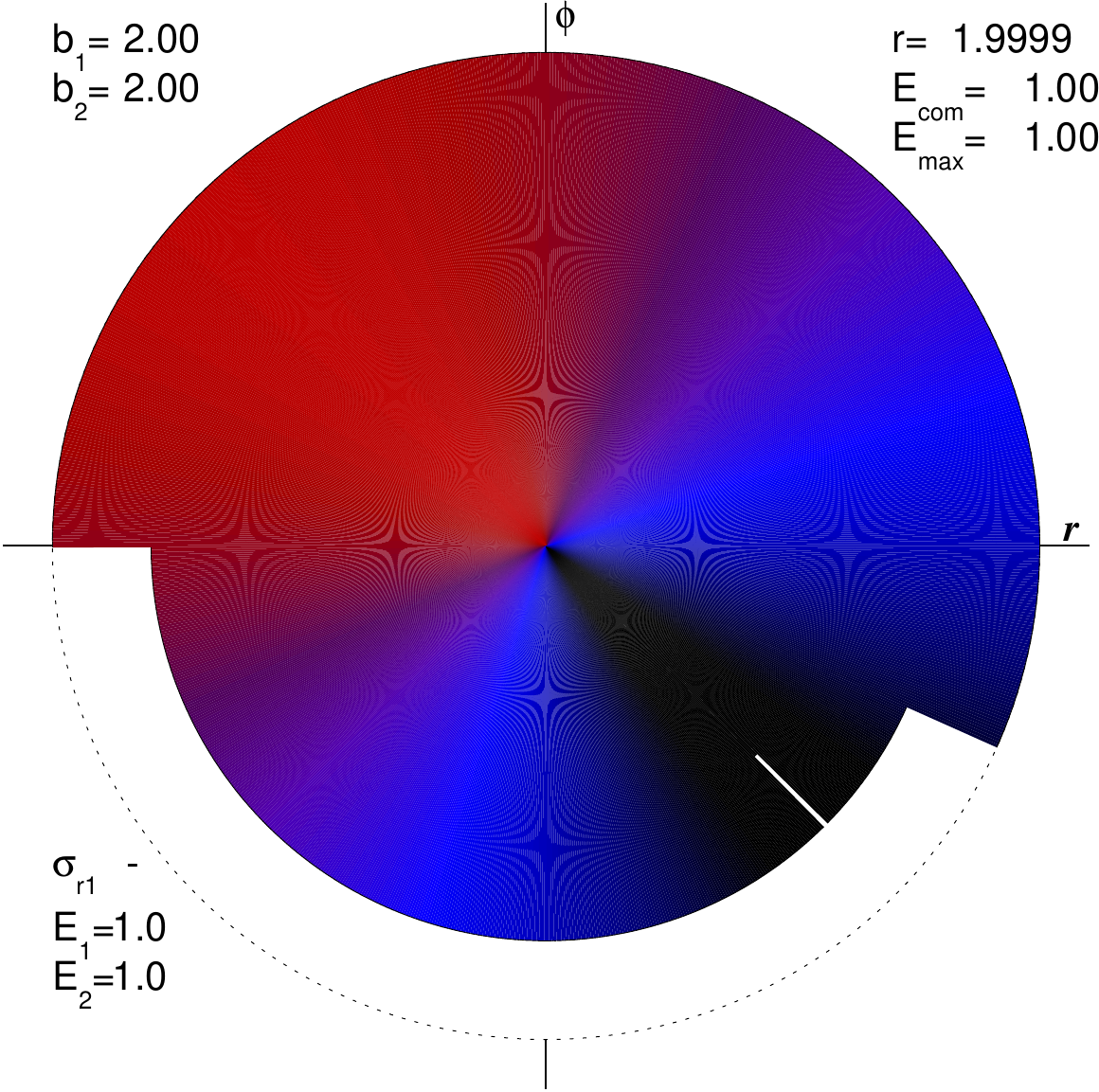}\hspace{2mm}
  \includegraphics[width=0.28\textwidth]{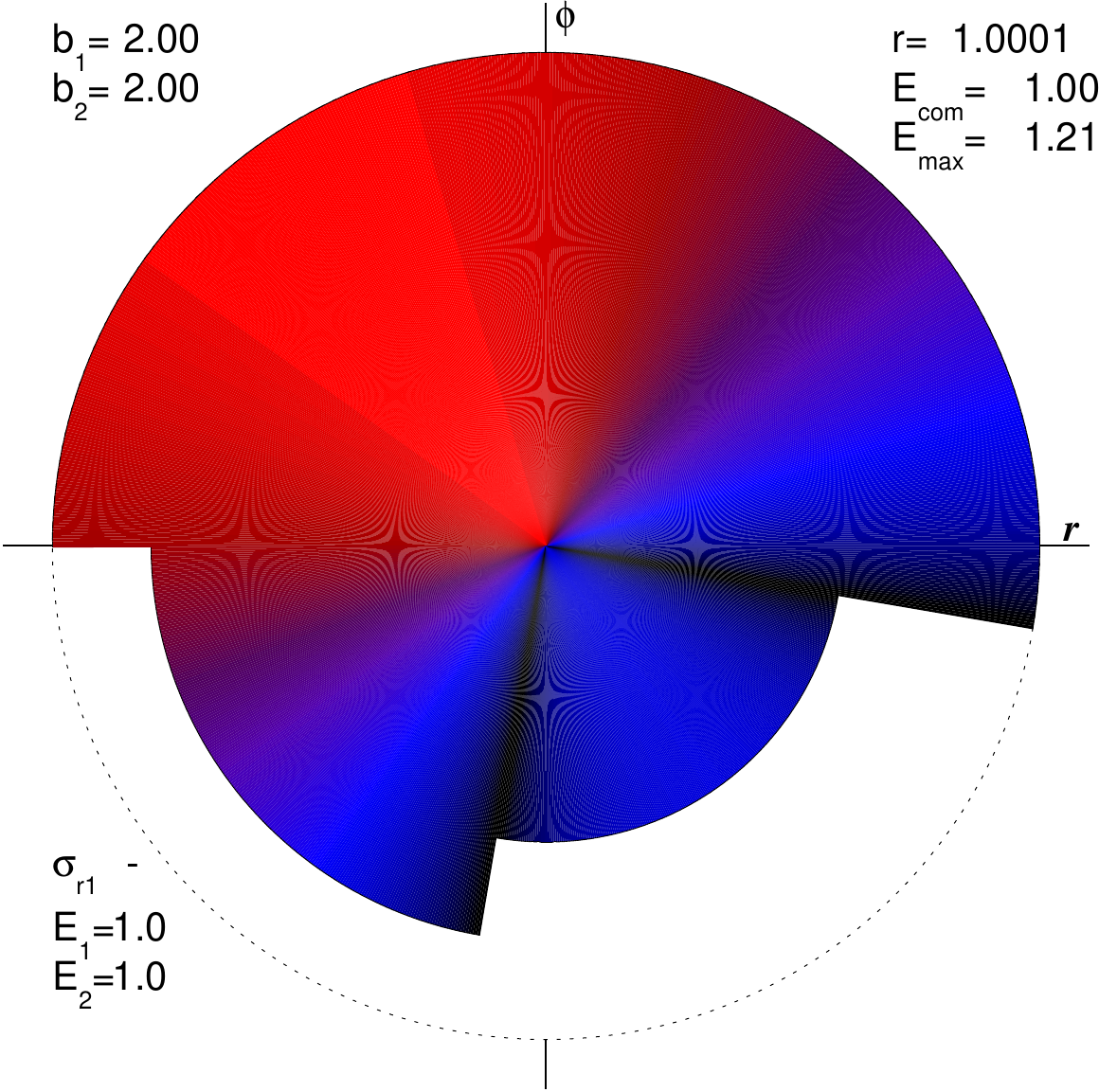}
\caption{\label{fig:Wald_polar} Polar plots of the energy and escape distribution of photons
  emitted in a classical Penrose decay process at various radii
  outside of an extremal Kerr black hole. The interpretation of
  these polar plots is described in detail in the text. The non-linear
  color scale represents the absolute value of the energy-at-infinity
  $|p_t^{(3)}|$ relative to the total input energy. $E_{\rm com}$ is
  the center of mass energy, normalized to the total energy of the
  incoming particles, and $E_{\rm max}$ is the maximum energy of the
  {\it escaping} particles, also normalized to total incoming energy.} 
\end{center}
\end{figure}

In Figure \ref{fig:Wald_polar} we show an example of this graphic for the
classical Penrose process of a single particle decay. Following our
approach in \cite{Schnittman:2018}, each image
corresponds to a specific choice of initial mass, energy, angular
momentum, and distance from the black hole. The polar coordinates are
defined in the particle's frame (or center-of-mass frame for
collisional reactions), with the coordinate radial direction oriented
to the right. The color represents the energy-at-infinity of the
daughter photons as a function of emission angle, and the radius of
the disk represents whether or not that photon escapes ($R=1$), is
captured by the black hole horizon ($R=0.8$), or has negative energy
($R=0.6$), in which case it will be also be captured by the black
hole. 

First of all, note that we are following the convention of Section
\ref{sec:BSW}, where we consider reactions of the form
\begin{equation}\label{eqn:p12_p34}
\mathbf{p}^{(1)}+\mathbf{p}^{(2)} \to
\mathbf{p}^{(3)}+\mathbf{p}^{(4)}\, .
\end{equation}
Particles 1 and 2 collide and produce particles 3 and 4. In the center
of mass frame, where all these polar plots are calculated,
particles 3 and 4 are always emitted in opposite directions. For the
classical Penrose decay process, we can still use the formalism of
equation (\ref{eqn:p12_p34}), with particles 1 and 2 having identical
trajectories and each one has exactly half of the total initial energy. In the first
frame of Figure \ref{fig:Wald_polar}, the decay takes place at
$r=10M$, relatively far from the black hole. The incoming particle is
falling from rest at infinity, and has the critical value of angular
momentum $b_1=2$. Thus the photons emitted in the forward-pointing,
$-\hat{r}$ direction have slightly higher energy due to Doppler
boosting. We can also see that roughly $20\%$ of all emitted photons
are captured by the black hole, preferentially those with negative
angular momentum. 

The middle frame of Figure \ref{fig:Wald_polar} corresponds to a decay
at $r=1.9999M$, just inside the ergosphere. At this point, we see the
first genuine Penrose process reaction, with the forward-going
particle having an energy just over unity, and the opposite particle
has a very slightly negative energy (the tiny notch in the polar plot
at $\phi \approx 315^\circ$). In the third frame, the reaction takes
place deep in the ergosphere, and the Wald limit is reached with
forward-pointing particles escaping with energy of $E_3 = 1.21E_1$. It
is interesting to note that, even this close to the event horizon,
when the initial particle has the critical value for angular momentum,
a majority ($\approx 53\%$) of the decay products are still able to
escape the black hole.

\begin{figure}
\begin{center}
  \includegraphics[width=0.5\textwidth]{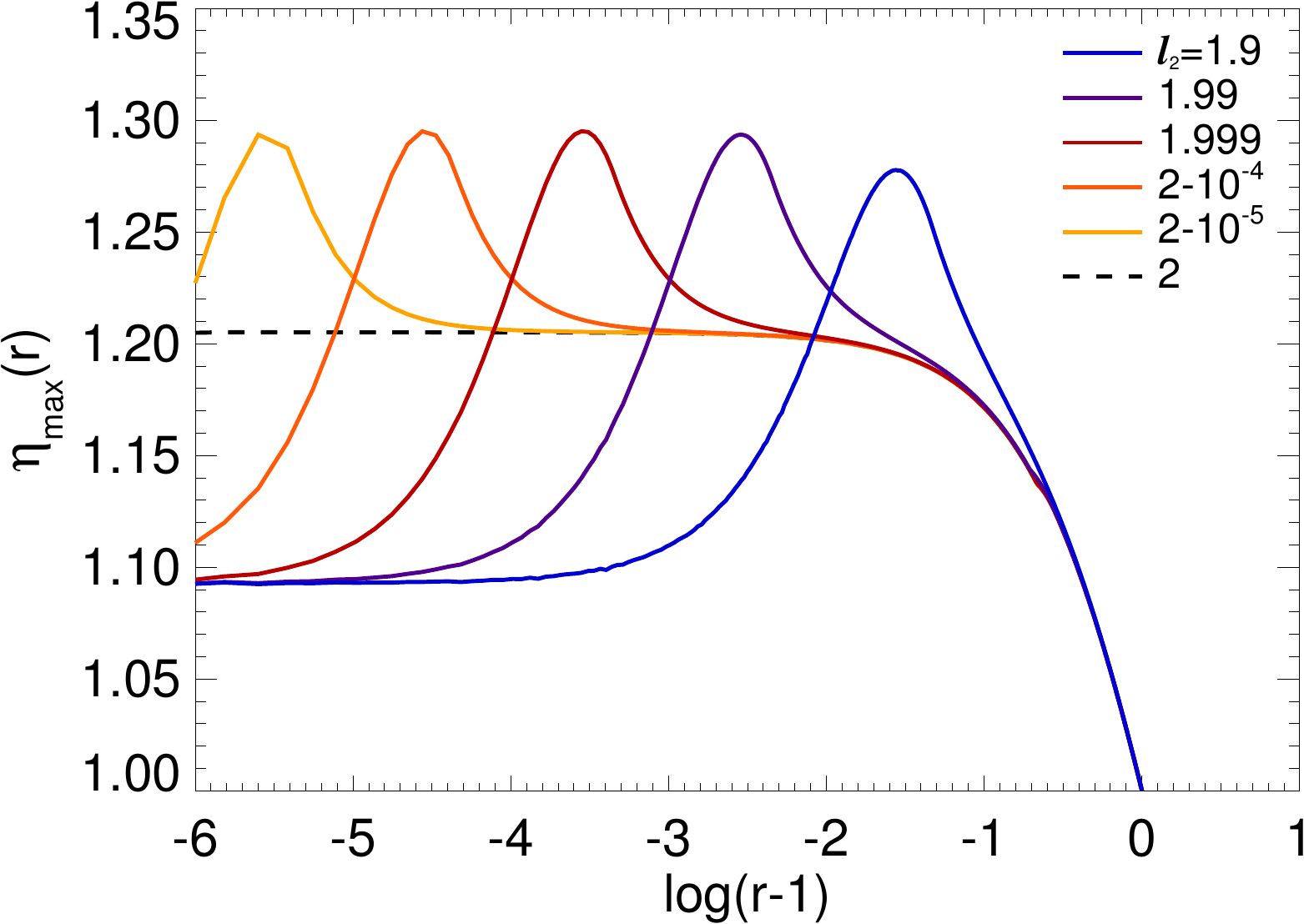}
\caption{\label{fig:BejgerFig2} 
  Peak efficiency for annihilation of equal-mass particles
  falling from rest at infinity, as a function of the radius at which
  the annihilation occurs. The angular momentum $b_1=2$ is fixed at
  the critical value, and $b_2$ varies. The black hole spin is
  maximal: $a=1$. (Compare to Fig.\ 2 of Ref.\ \cite{Bejger:2012})}
\end{center}
\end{figure}

Next we turn to the collisional cases, starting with the analysis of
Bejger {\it et al.} \cite{Bejger:2012}. As mentioned briefly above,
they focus on the energy of the particles that can eventually escape
to infinity. They also restrict their considerations to identical, massive
particles falling from rest at infinity ($E_1=E_2=1$) and particle one
has a critical value of angular momentum $b_1=2$. Both daughter
particles are massless, so we also refer to the process as
annihilation. In Figure
\ref{fig:BejgerFig2} we show the peak energy for $E_3$ as a function of
reaction radius for a range of select values of $b_2$. The peak
quickly asymptotes to $\sim 130\%$, only marginally higher than that
of a single particle decay. Ref.\ \cite{Leiderschneider:2016} derives
this efficiency analytically as
$\eta=(1+\sqrt{3}+\sqrt{6})/4$. 

\begin{figure}
\begin{center}
  \includegraphics[width=0.07\textwidth]{pf00.pdf}\hspace{2mm}
  \includegraphics[width=0.28\textwidth]{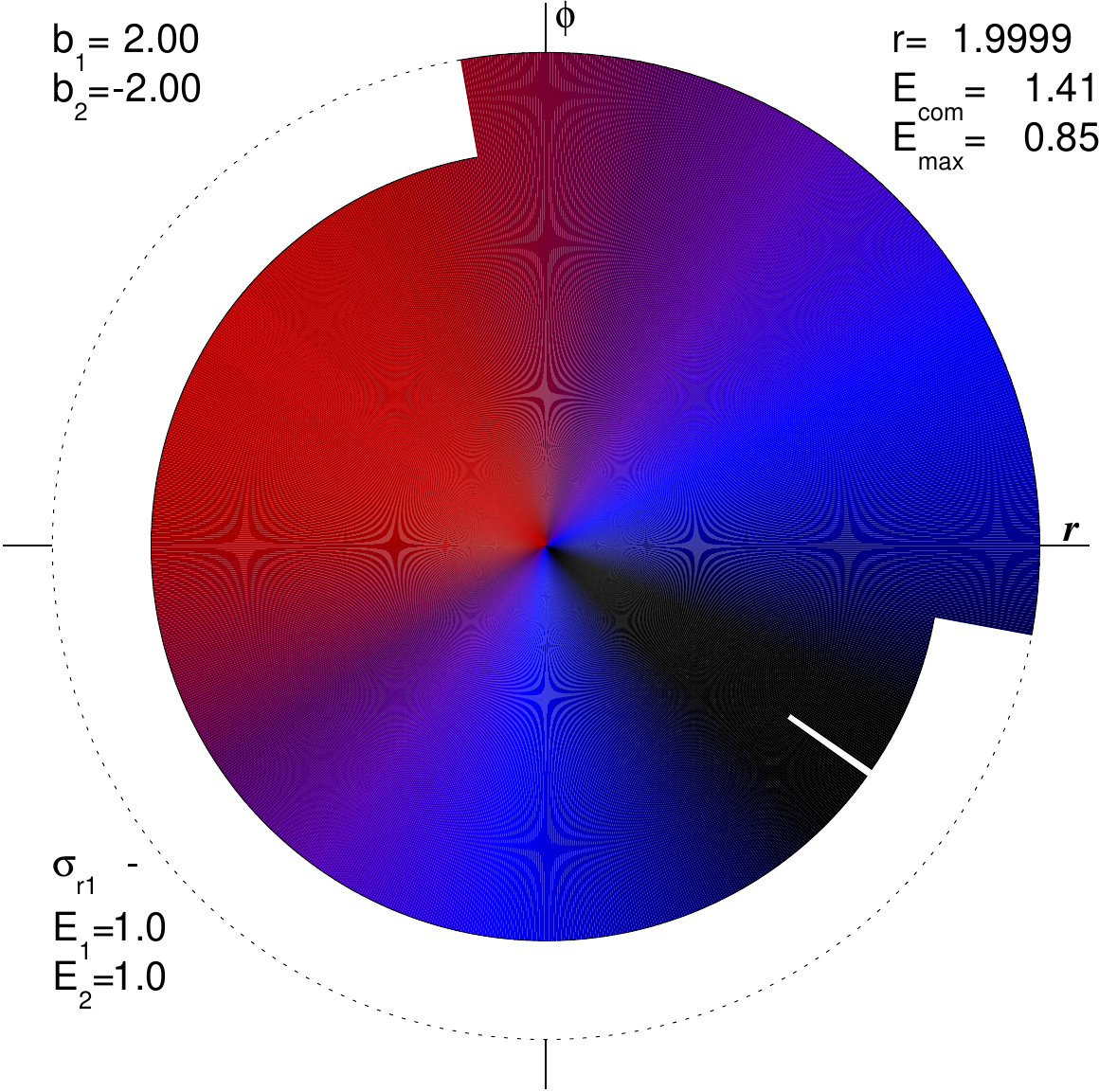}\hspace{2mm}
  \includegraphics[width=0.28\textwidth]{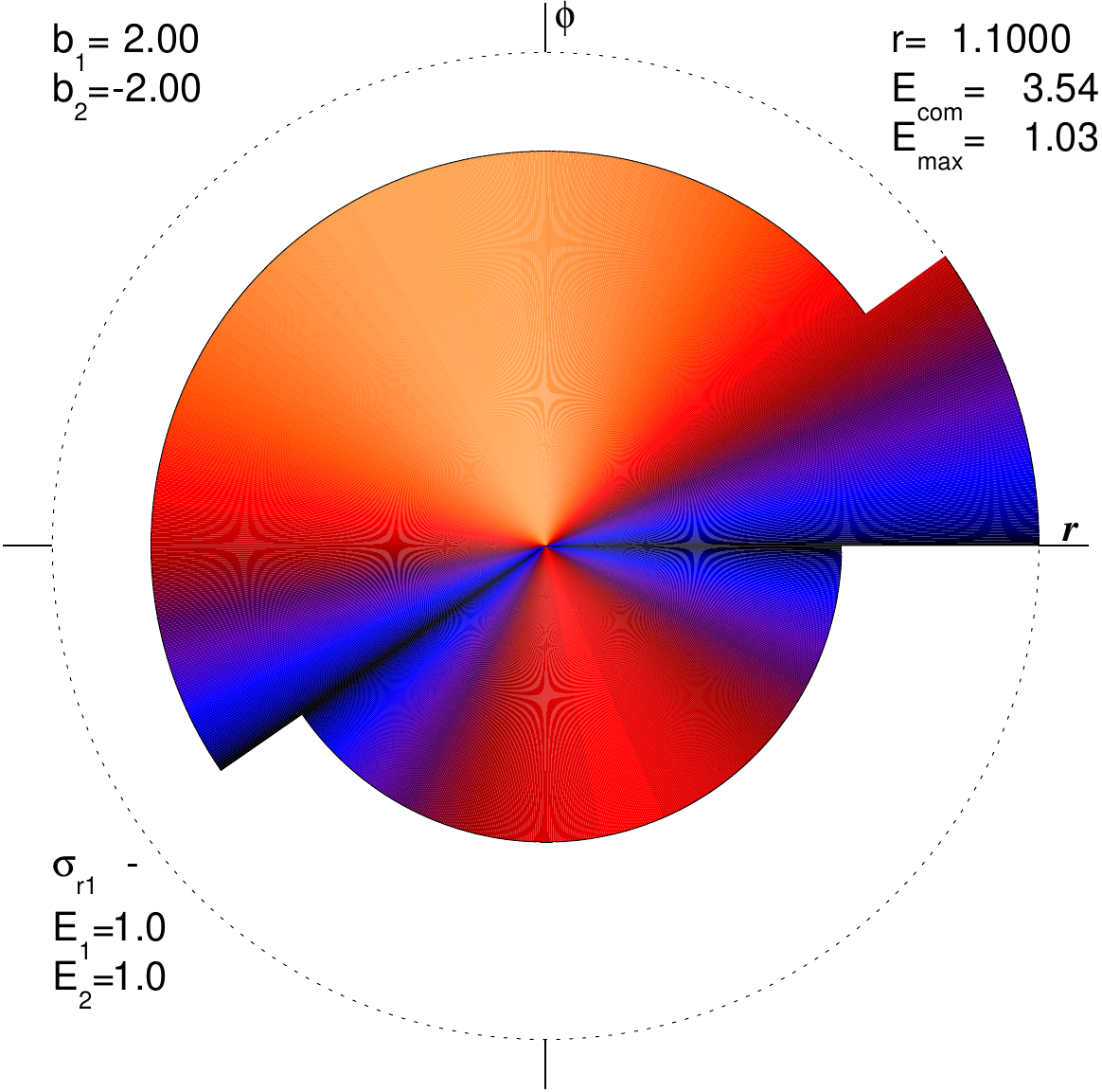}\hspace{2mm}
  \includegraphics[width=0.28\textwidth]{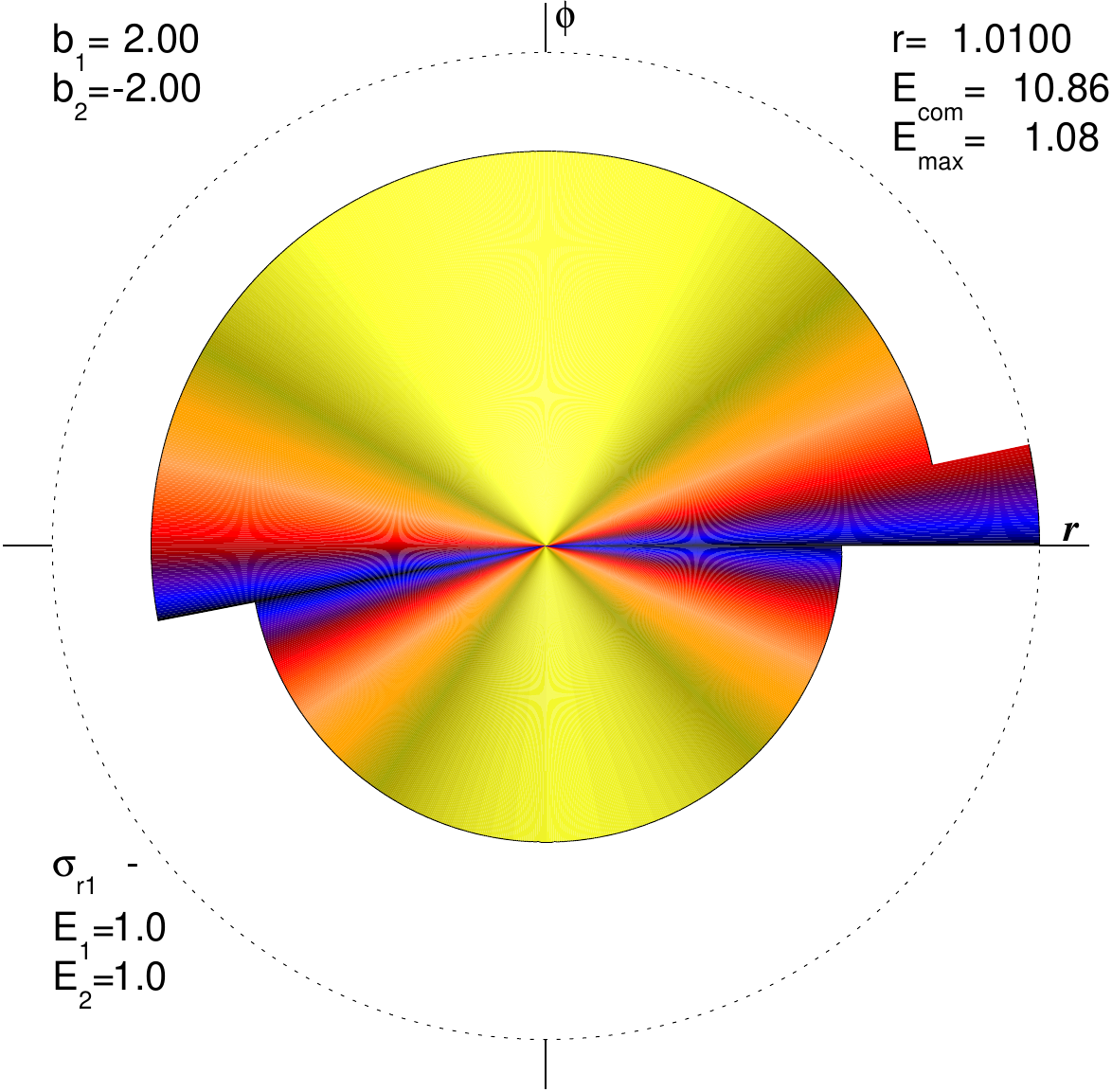}
\caption{\label{fig:Bejger_polar}
  Polar plots of the energy and escape distribution of photons
  emitted in a Penrose annihilation reaction at various radii
  outside of an extremal Kerr black hole.}
\end{center}
\end{figure}

In Figure \ref{fig:Bejger_polar} we show similar results, but with our
polar energy plots for a range of annihilation radii, while keeping
the initial particle properties fixed: $b_1=2$ and $b_2=-2$. In all
cases, both particles are on 
inward-moving trajectories $p_r <0$, denoted in the figure with a '-'
sign for the parameter $\sigma_{r1}$. While the
c.o.m. energy grows with decreasing radius, the escape fraction is
significantly reduced, and most of the high-energy particles are
captured by the black hole. As described in \cite{Bejger:2012}, even
the escaping particles are actually initially emitted with $p_r<0$,
but reflect off of the centrifugal barrier of the black hole before
escaping to infinity. This is not entirely obvious in Figure
\ref{fig:Bejger_polar}, which shows the angular distribution as
measured in the particle's center-of-mass frame. Deep into the
particles' plunge, even the particles emitted in the $+r$
direction of the polar plots in fact have inward-moving trajectories
in the coordinate frame.

Completely independent, and largely ignorant, of the flurry of papers
surrounding BSW, at that time we were working on calculating the phase
space distribution and annihilation rates of dark matter particles
around a spinning black hole \cite{Schnittman:2015}. Adhering to the
well-known strategy of ``if your only tool is a 
hammer, everything looks like a nail,'' we developed a version of the
\pan\, ray-tracing code \cite{Schnittman:2013} to calculate fully 3-dimensional
trajectories of massive test particles coming in from rest at
infinity. A sample of these particles will annihilate, and then we
follow the photon trajectories either to the horizon or escape to
infinity. 

Curiously, some of these escaping photons would have very large
energy, ten times greater than the rest mass of the dark 
matter particles, in clear contradiction to the analytic predictions
of \cite{Bejger:2012,Harada:2012}. After months of searching for bugs
and mathematical errors, we were forced to accept the results as
physically real, and were then able to isolate and identify the cause
of the discrepancy with the analytic results. The difference was
really quite simple: all previous studies had limited their attention
to incoming particles with critical values for $b_1=2$, allowing the
particles to get as close as possible to the event horizon in order to
maximize the center-of-mass energy. Yet the numerical approach
naturally included a much greater sample of phase space, including
particles with $b_1>2$ that reflected off the centrifugal barrier
before colliding with other incoming particles. It was these
out-going, super-critical particles responsible for the escaping
high-energy annihilation products, and thus we call them {\it
  super-Penrose} processes. 

\begin{figure}
\begin{center}
  \includegraphics[width=0.5\textwidth]{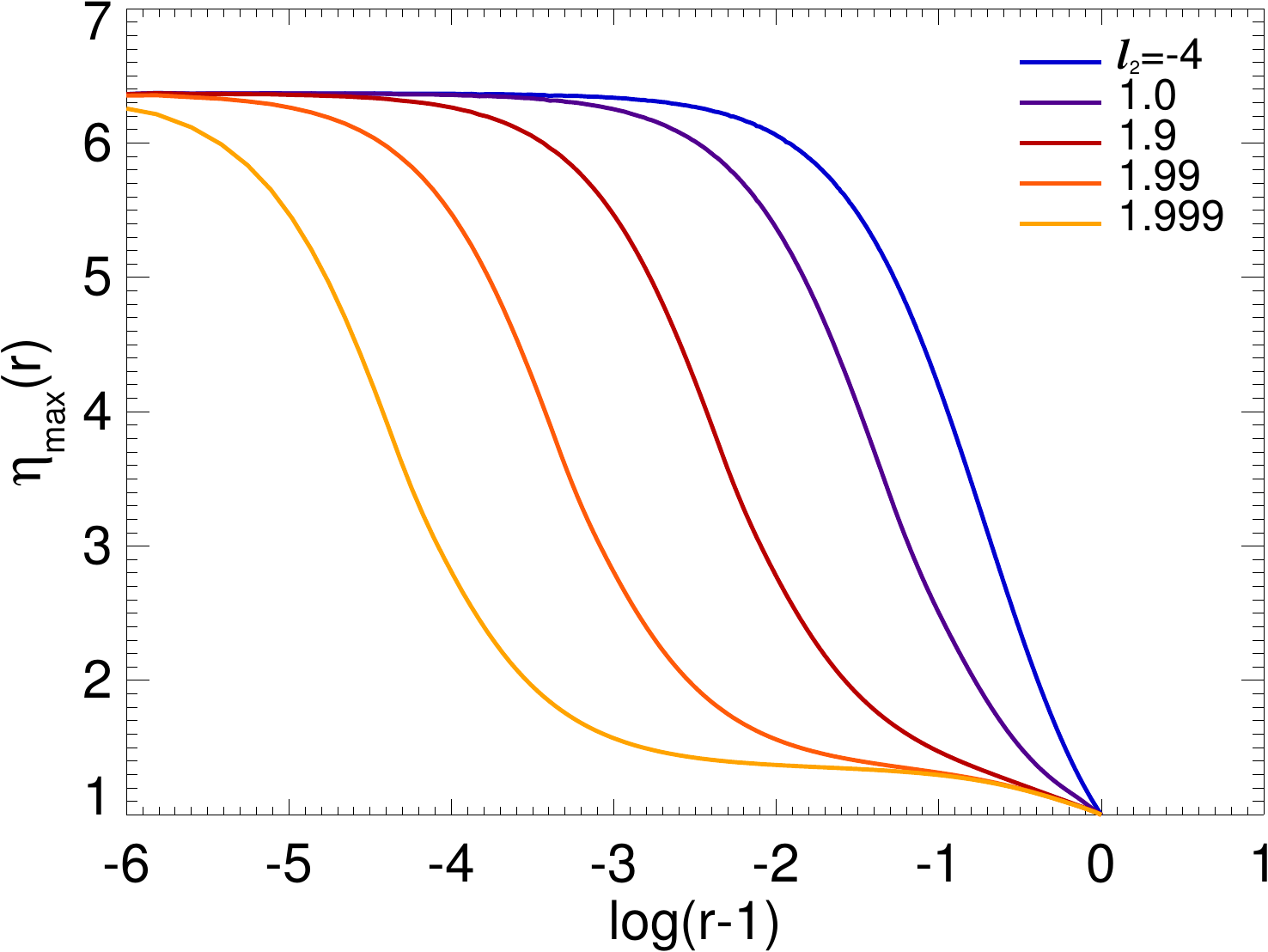}
\caption{\label{fig:SchnittmanFig3}
  Peak efficiency for annihilation of equal-mass particles
  falling from rest at infinity, as a function of the radius at which
  the annihilation occurs. Unlike in Fig.\ \ref{fig:BejgerFig2}, here we
  allow $p_r^{(1)}>0$, which greatly increases the fraction and energy
  of escaping photons. The angular momentum $b_1=2$ is fixed at
  the critical value, and $b_2$ varies. The black hole spin is
  maximal: $a=1$.} 
\end{center}
\end{figure}

As with the ingoing trajectories, the c.o.m. energy of the outgoing
particles also is maximized near the horizon, so we want to focus on
near-critical particles with $b_1\approx 2$. We can now return to
an analytic approach, simply changing the sign for the radial momentum
in equation (\ref{eqn:p_r}). The results are plotted
in Figure \ref{fig:SchnittmanFig3}, showing the peak efficiency as a
function of radius for a selection of $b_2$ values. Recall our
definition of efficiency is ``total energy out divided by total energy
in,'' so a photon that escapes with $\eta=6.37$ actually has an energy
of $\approx 13\times m_1$, in agreement with the results found
accidentally in \cite{Schnittman:2015}. As with the ingoing
annihilation, an analytic expression was derived in
\cite{Leiderschneider:2016}:
$\eta_{\max}=(2+\sqrt{3})(2+\sqrt{2})/2$. 

\begin{figure}
\begin{center}
  \includegraphics[width=0.07\textwidth]{pf00.pdf}\hspace{2mm}
  \includegraphics[width=0.28\textwidth]{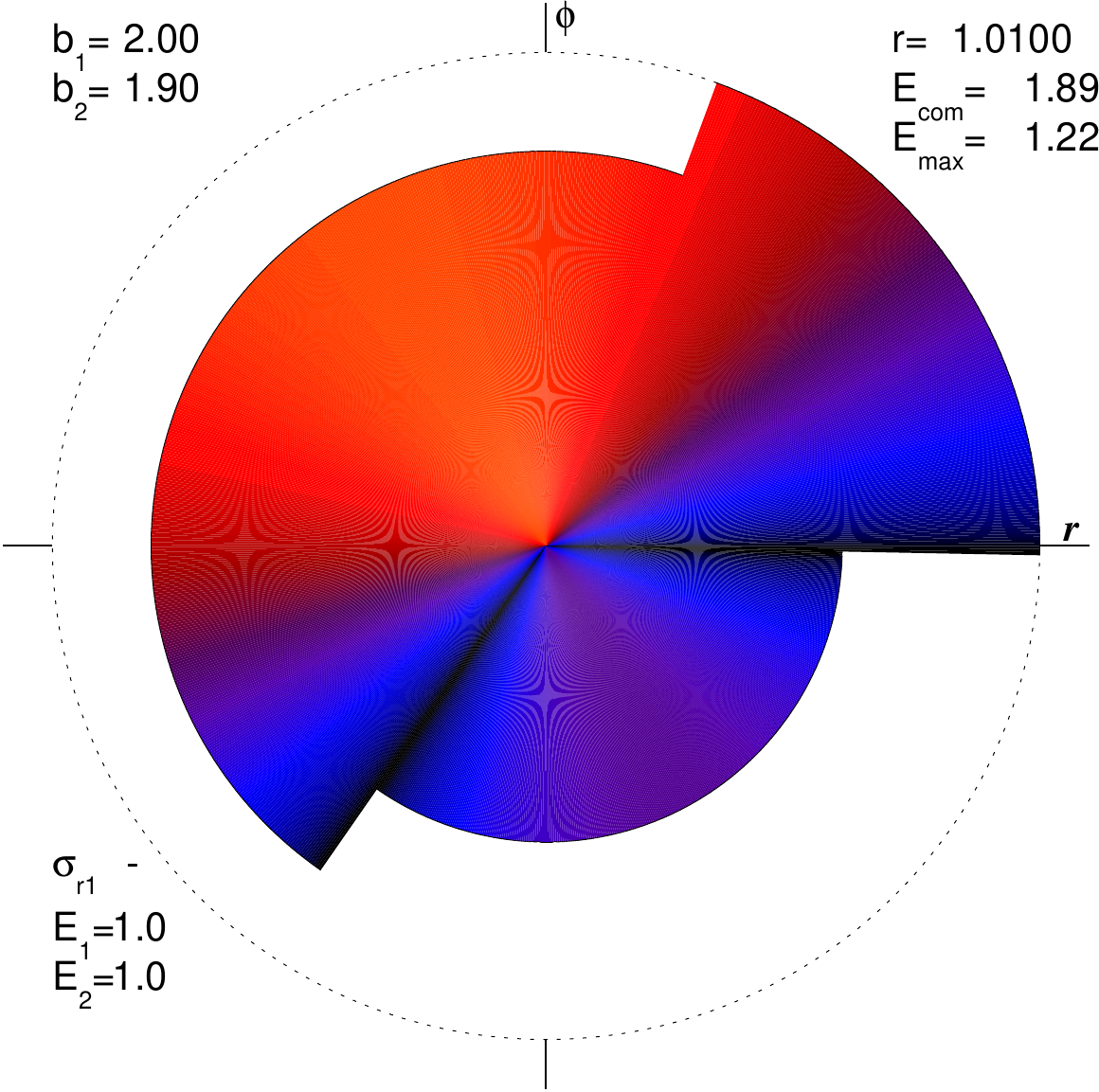}\hspace{2mm}
  \includegraphics[width=0.28\textwidth]{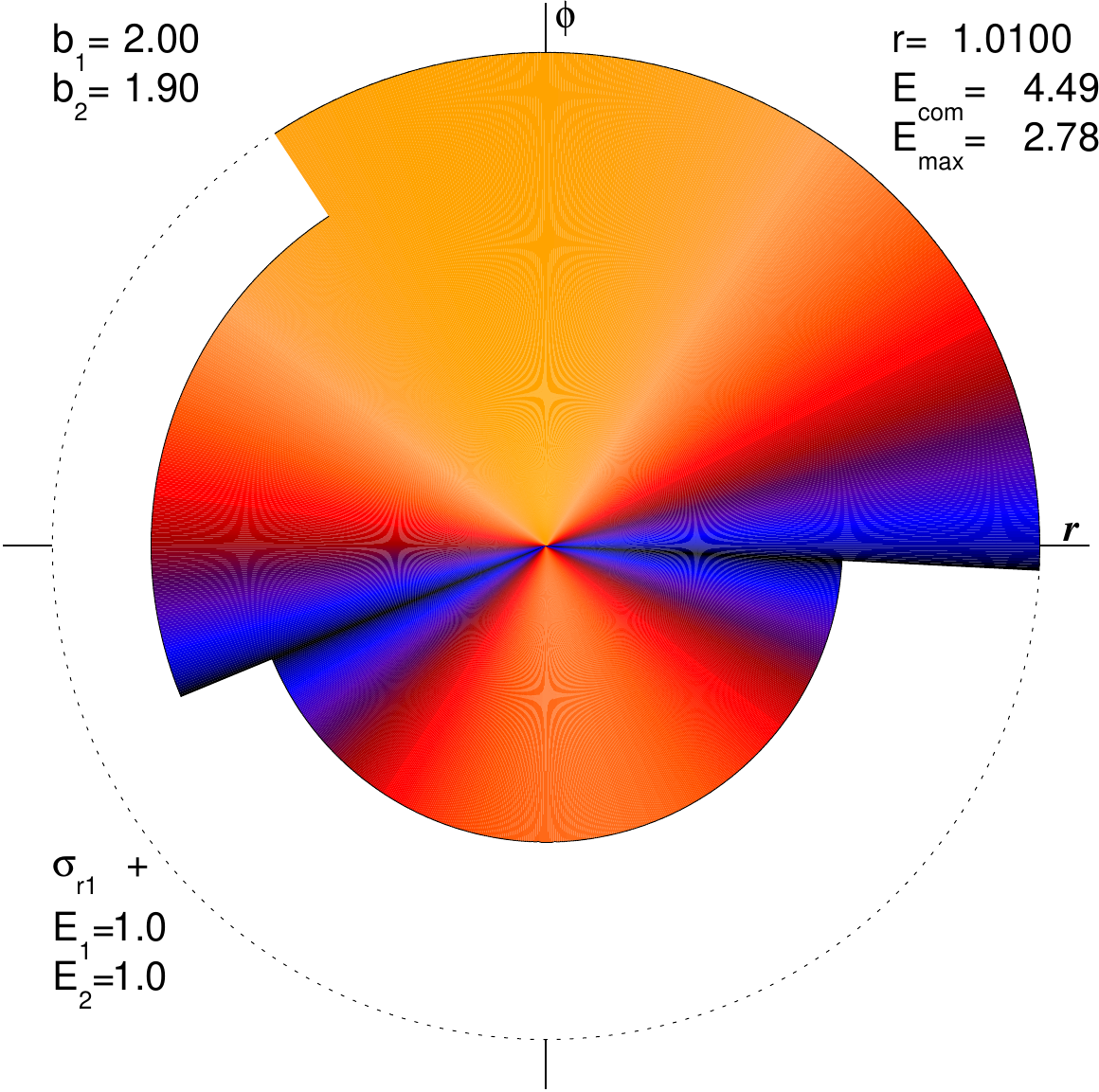}
\caption{\label{fig:polar_plusminus}
  Polar plots of the energy and escape distribution of photons
  emitted in a Penrose annihilation reaction at $r=1.01M$
  outside of an extremal Kerr black hole. All parameters are
  identical, except the left-hand plot is for two ingoing particles,
  and the right-hand plot has an outgoing particle (1).}
\end{center}
\end{figure}

The difference between $p_r^{(1)} >0$ and $p_r^{(1)}<0$ is also shown
in the polar plots of Figure \ref{fig:polar_plusminus}, corresponding
to the blue curve in Figure \ref{fig:BejgerFig2} and the dark red
curve of Figure \ref{fig:SchnittmanFig3}. All parameters are
identical, except one has an outgoing particle 1. This small detail
makes a very big difference in the center of mass energy
[$1.89(E_1+E_2)$ vs $4.49(E_1+E_2)$], escape fraction ($20\%$ vs
$35\%$), and peak efficiency ($122\%$ vs $278\%$). Note that the other
parameters, in particular $b_2$ and $r$, are not specifically
chosen to optimize energy efficiency, but rather as a representative
sample of the literature. 

\begin{figure}
\begin{center}
  \includegraphics[width=0.07\textwidth]{pf00.pdf}\hspace{2mm}
  \includegraphics[width=0.28\textwidth]{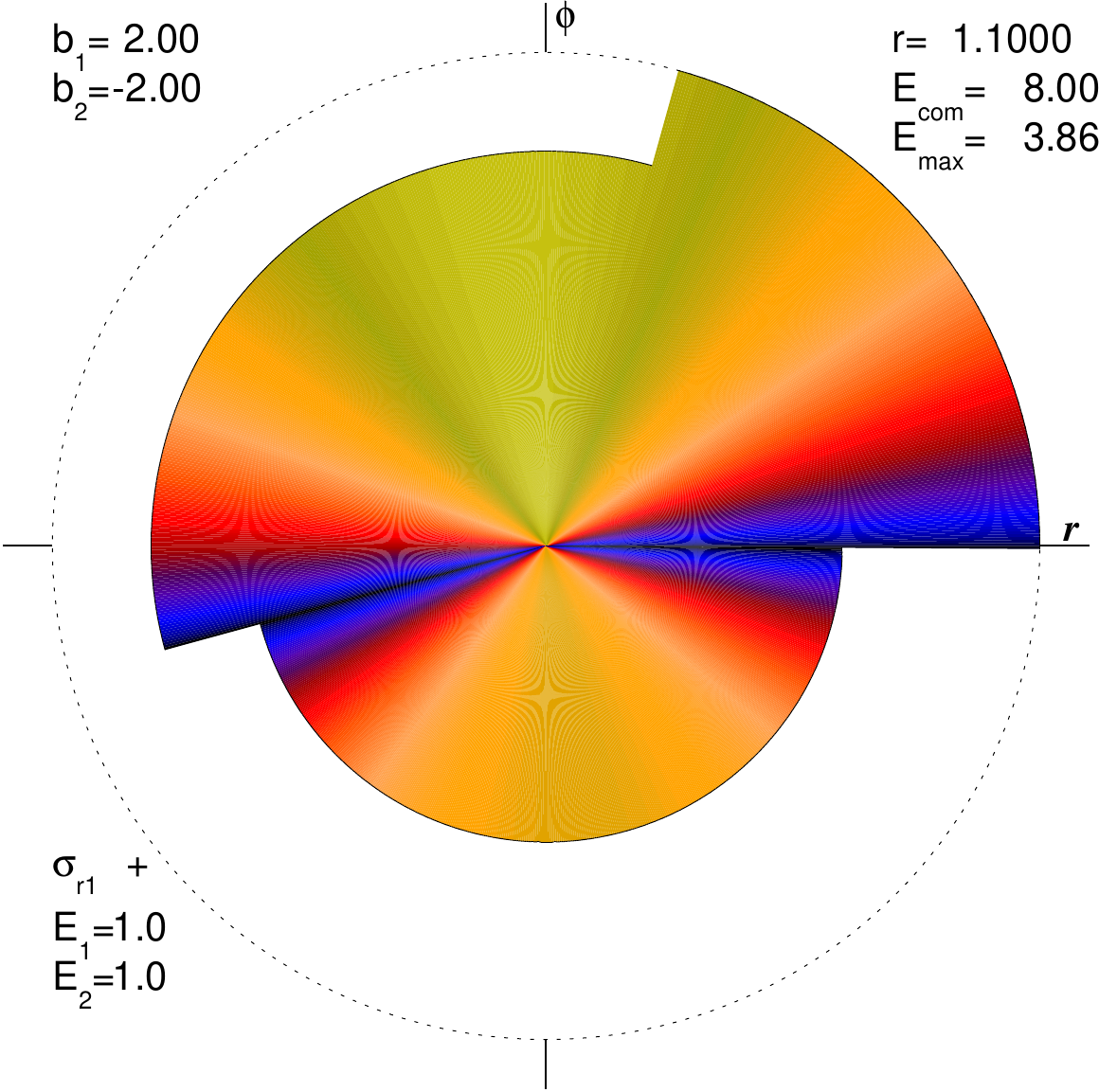}\hspace{2mm}
  \includegraphics[width=0.28\textwidth]{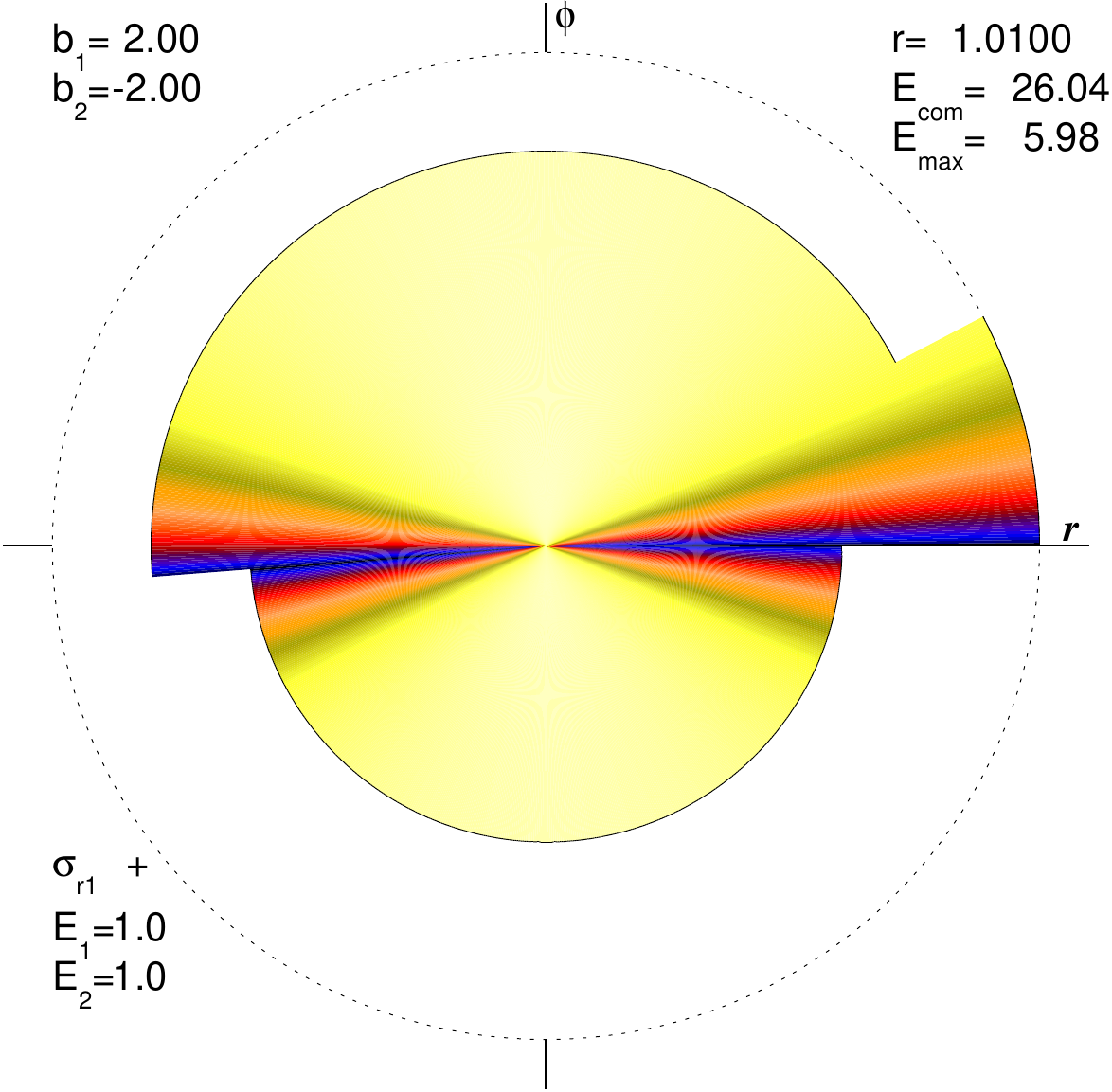}\hspace{2mm}
  \includegraphics[width=0.28\textwidth]{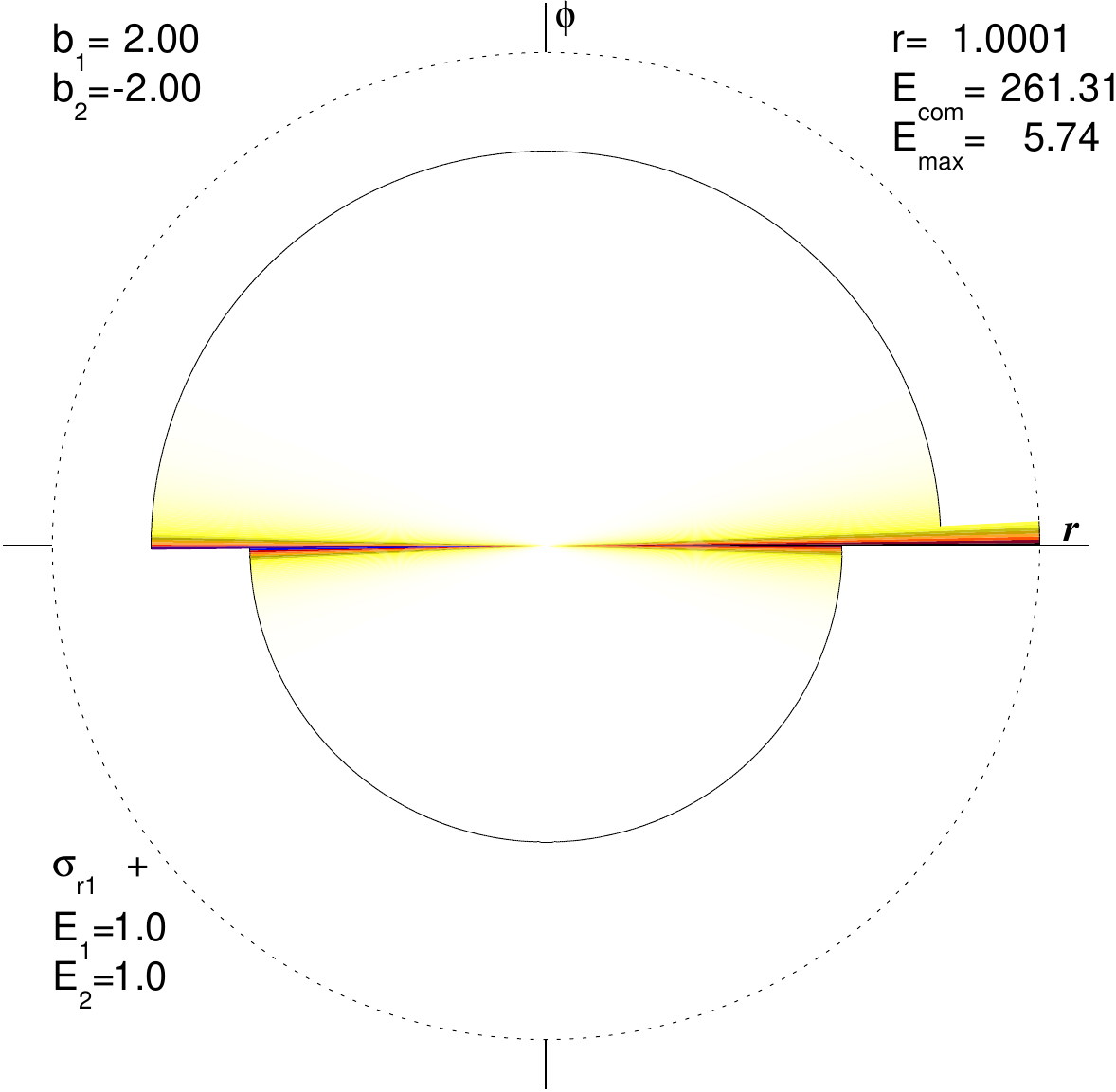}
\caption{\label{fig:polar_eff_r}
  Polar plots of the energy and escape distribution of photons
  emitted in a Penrose annihilation reaction for nearly head-on collisions
  outside of an extremal Kerr black hole. As the collision radius
  approaches the horizon, the c.o.m. energy diverges while the escape
  fraction approaches zero.}
\end{center}
\end{figure}

In Figure \ref{fig:polar_eff_r} we show the energy and escape
distributions for collisions tuned to maximize efficiency. The choice
of $b_2=-2$ leads to more head-on collisions, and nicely demonstrates
the nearly symmetric distribution of annihilation product energies due
to the zero net angular momentum of the initial particles [$b_2 =
-2(1+\sqrt{2})$ gives an even greater energy, but is less symmetric
and has a slightly smaller escape fraction]. 

\begin{figure}
\begin{center}
  \includegraphics[width=0.07\textwidth]{pf00.pdf}\hspace{2mm}
  \includegraphics[width=0.28\textwidth]{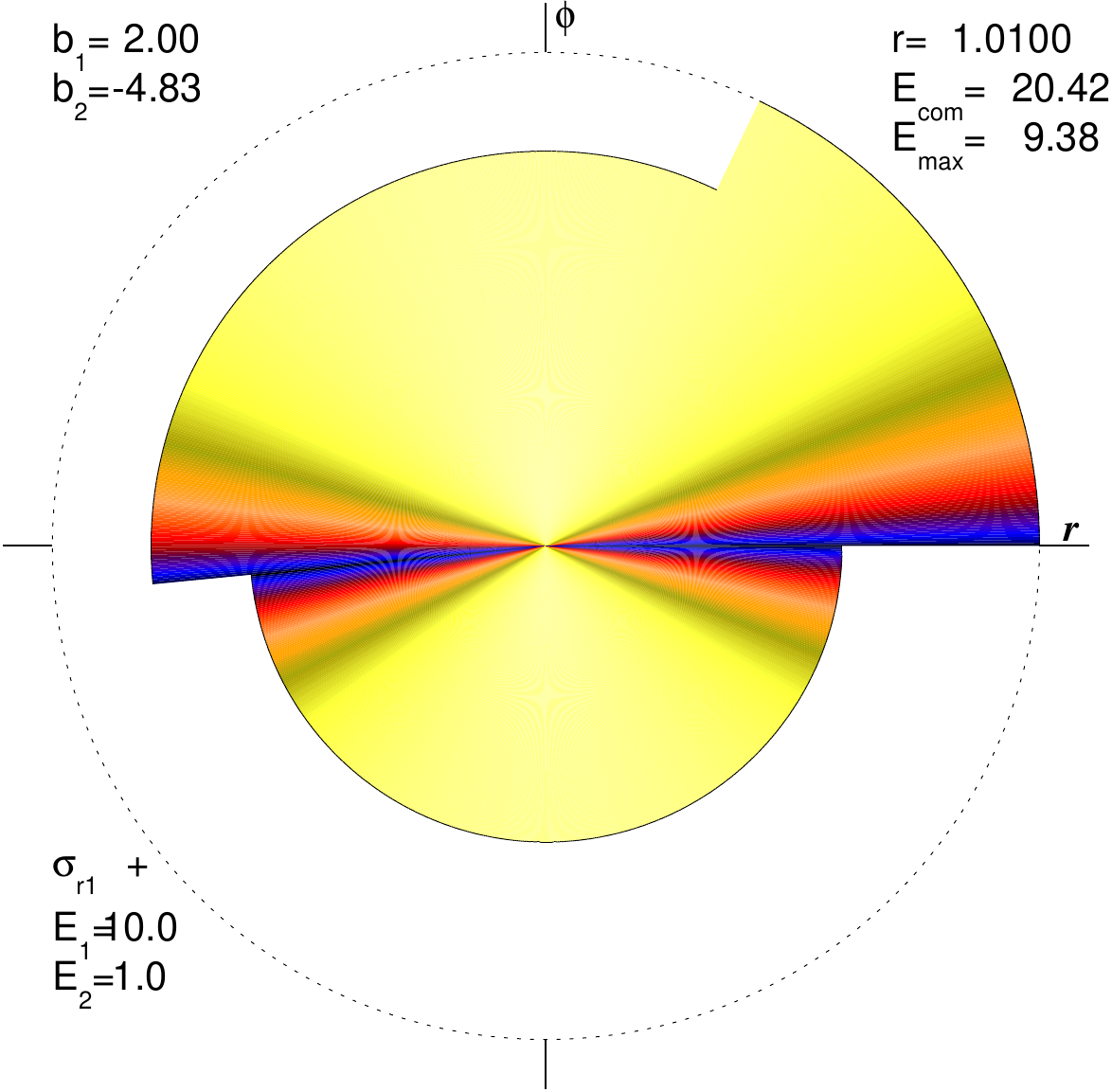}\hspace{2mm}
  \includegraphics[width=0.28\textwidth]{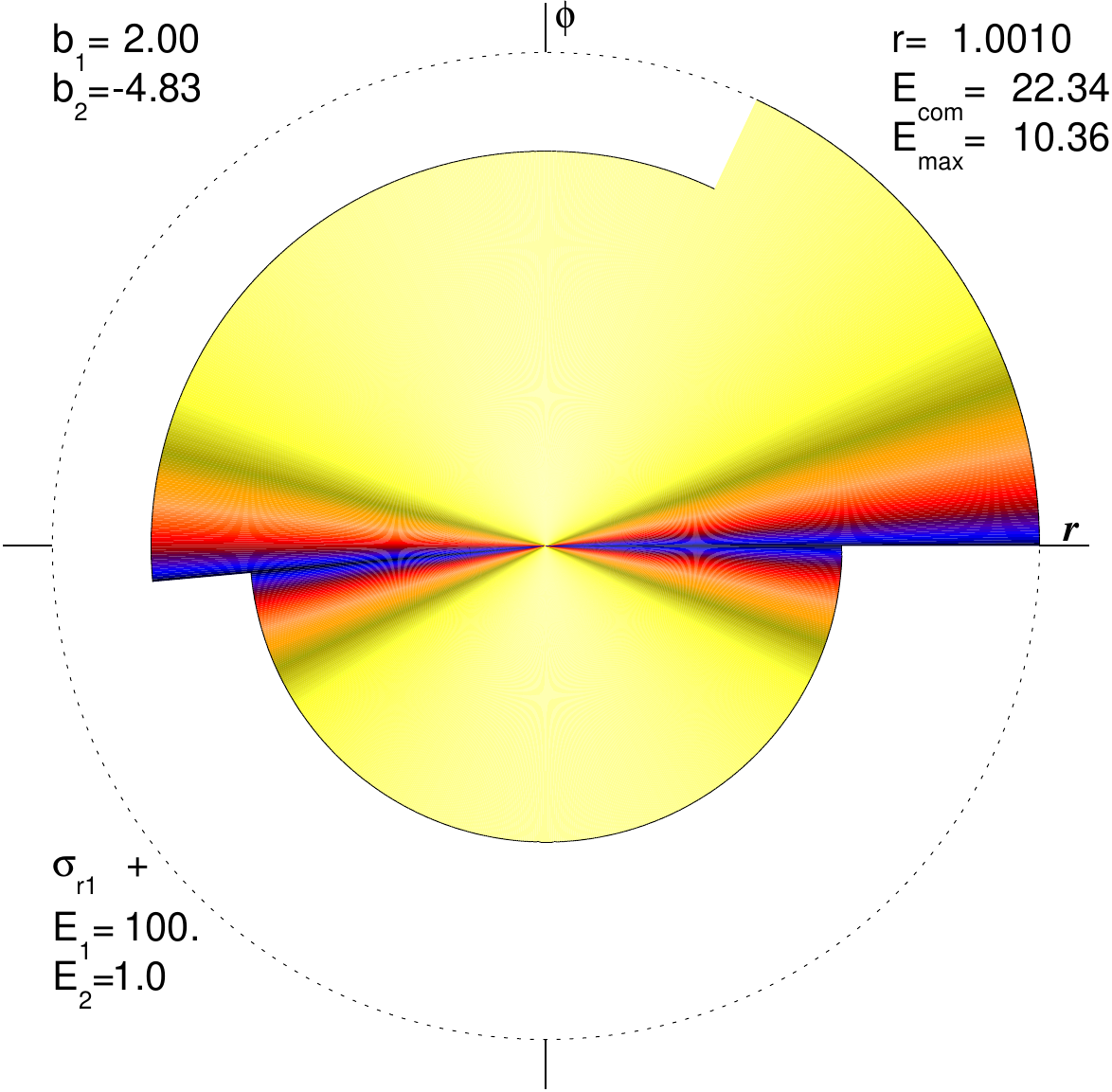}\hspace{2mm}
  \includegraphics[width=0.28\textwidth]{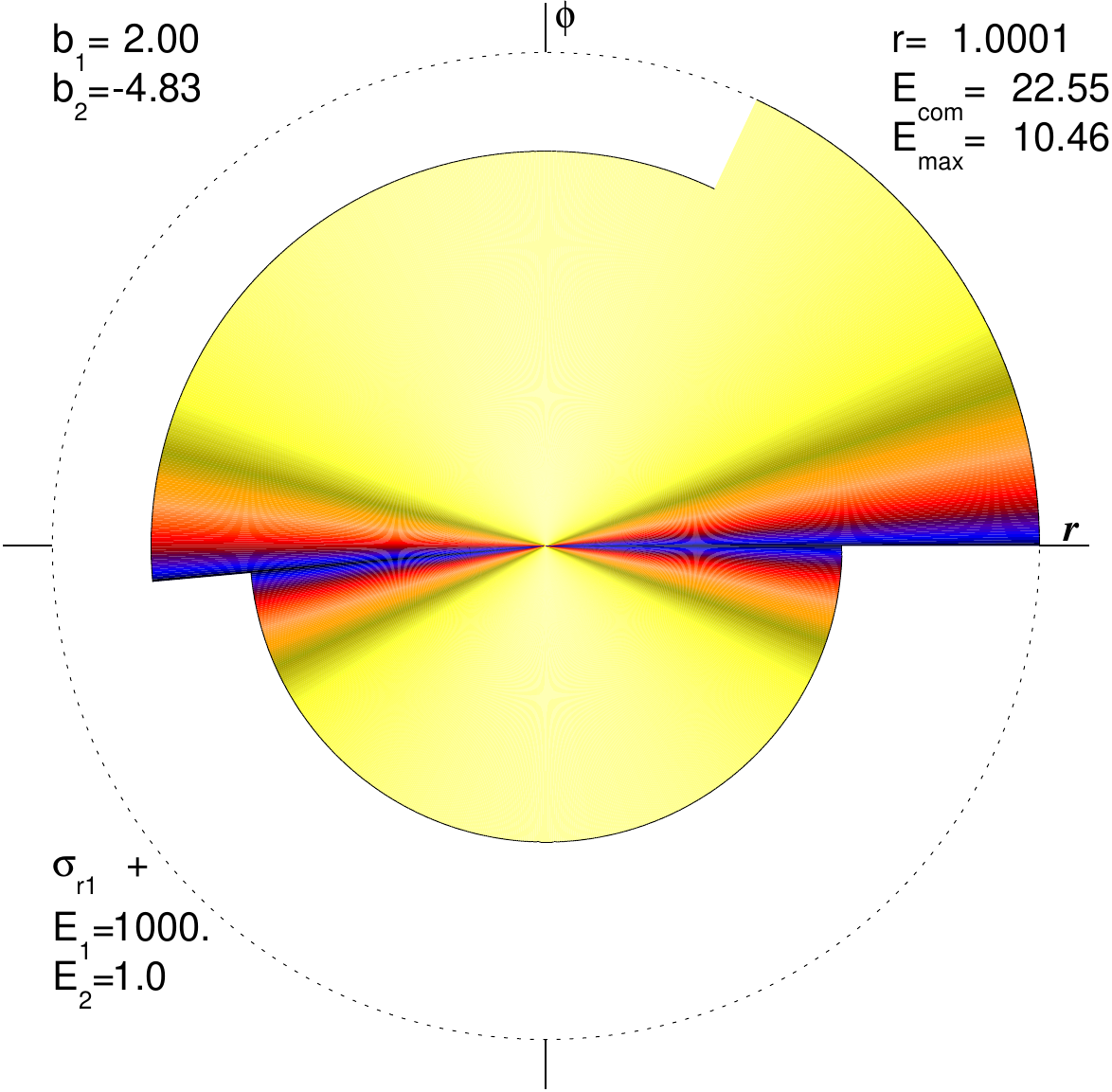}
\caption{\label{fig:polar_compton}
  Polar plots of the energy and escape distribution of photons
  resulting from a Compton-like scattering event
  outside of an extremal Kerr black hole. The initial particles
  include a photon and massive particle, both falling in from
  infinity. As the energy of the photon increases, we need to move the
  position of the collision closer and closer to the horizon in order
  to achieve the target efficiency of $1000\%$.}
\end{center}
\end{figure}

We showed in \cite{Schnittman:2014} that the absolute maximum
efficiency is achieved for Compton-like scattering between an
outgoing photon with $b_1=2$ and an infalling massive particle with
$b_2=-2(1+\sqrt{2})$. The post-scatter products are an in-going photon
with $b_3=2$ and an in-going massive particle with negative energy and
angular momentum. Figure \ref{fig:polar_compton} shows the energy and
escape distributions for these Compton scattering reactions, for a
range of photon energy $E_1$. In the limit of $E_1 >> E_2$ and $r\to
r_+$, the absolute maximum efficiency for Compton scattering is given
by $\eta = (2+\sqrt{3})^2\approx 1392\%$ \cite{Leiderschneider:2016}!

The most remarkable feature of this
particular configuration is that, after the scattering event, the
photon---now boosted by in energy by a factor of $\gtrsim 10$---is on an
in-going trajectory, reflects off the black hole's centrifugal
barrier, and then becomes an out-going photon. At this point, it can
scatter off a new infalling massive particle. In this way, the panels
of Figure \ref{fig:polar_compton} can be considered as three
consecutive scattering events, each photon getting boosted to a higher
energy, while the massive particles all have the same basic
energy. 

By including these multiple scattering events, the net efficiency can
grow without limit. Well, almost without limit. We also showed in
\cite{Schnittman:2014} that each step in this scattering process
deposits more negative energy and angular momentum into the black
hole, resulting in its eventual spin-down. Recall from Section
\ref{sec:BSW} above, for an incremental decrease in spin $\epsilon$,
the critical impact parameter for reflection increases to $b_{\rm
  crit} \approx 2(1+\epsilon^{1/2})$, steadily pushing the location
for high-efficiency collisions farther from the horizon. Taking each
scattering event as 
an increase in photon energy of a factor of 10, the spin parameter
after $N$ scatters is given by \cite{Schnittman:2014}
\begin{equation}
1-a_\ast=\epsilon_{N+1} \approx (4+2\sqrt{2})N\frac{m_2}{M}\, .
\end{equation}

The requirement that the scattering event occurs outside of $r_{\rm
  crit}$ leads to a condition on the maximum number of such events to
be $N_{\rm max} \approx \log_{10}(M/m_2)^{1/2}$ (note also in Figure
\ref{fig:polar_compton} how each higher energy requires collision at a
smaller radius in order to achieve the target $10\times$ increase in
energy). Taking $m_2$ to be
the electron mass and $M=10M_\odot$ gives $N_{\rm max} \approx 30$,
for a peak energy of $10^{26}$ GeV. Thus 
photons undergoing repeated Compton scattering events could not only
far surpass the Planck energy scale, but these hyper-energetic photons
could even escape to an observer at infinity! Furthermore, by
accelerating the photons to high energy one step at a time, it avoids
the BSW limit derived above of $\epsilon \lesssim 10^{-76}$ to a much
larger $\epsilon \lesssim 10^{-59}$. Unfortunately, this is still too
small for an astrophysical black hole to attain due to the rapid
accretion of CMB photons and neutrinos, but perhaps might still be
attainable in a properly shielded laboratory environment.

Following shortly after the discovery of these super-Penrose
reactions, Berti {\it et al.} \cite{Berti:2015} found solutions with
even greater efficiencies (aptly named ``super-Schnittman''
reactions), but only for trajectories that were not
obtainable with particles falling in from infinity. Specifically, they
consider very similar configurations to the Compton scattering
described above, with $p_r^{(1)}>0$ and $p_r^{(2)}<0$, but now with
$b_1<2$, which is only possible if particle 1 originates in the
ergosphere via some other scattering process
\cite{Berti:2015,Zaslavskii:2016}. The
greatest efficiency is found for the greatest deviation from the
critical value of $b_{\rm crit}=2$. Unfortunately, the larger the
deviation, the harder it is to produce such a particle by colliding
``normal'' particles falling in from infinity. 

\begin{figure}
\begin{center}
  \includegraphics[width=0.07\textwidth]{pf00.pdf}\hspace{2mm}
  \includegraphics[width=0.28\textwidth]{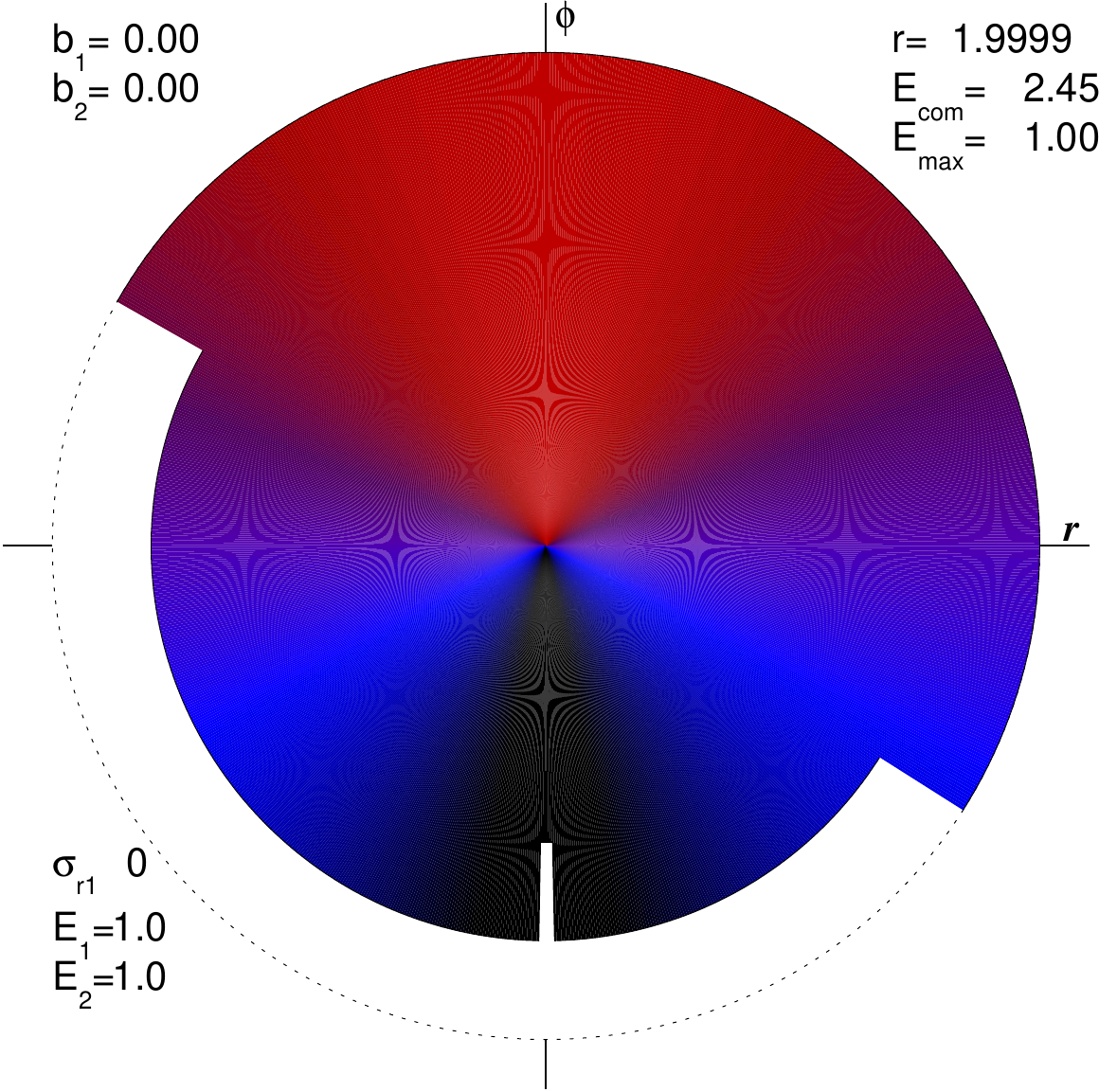}\hspace{2mm}
  \includegraphics[width=0.28\textwidth]{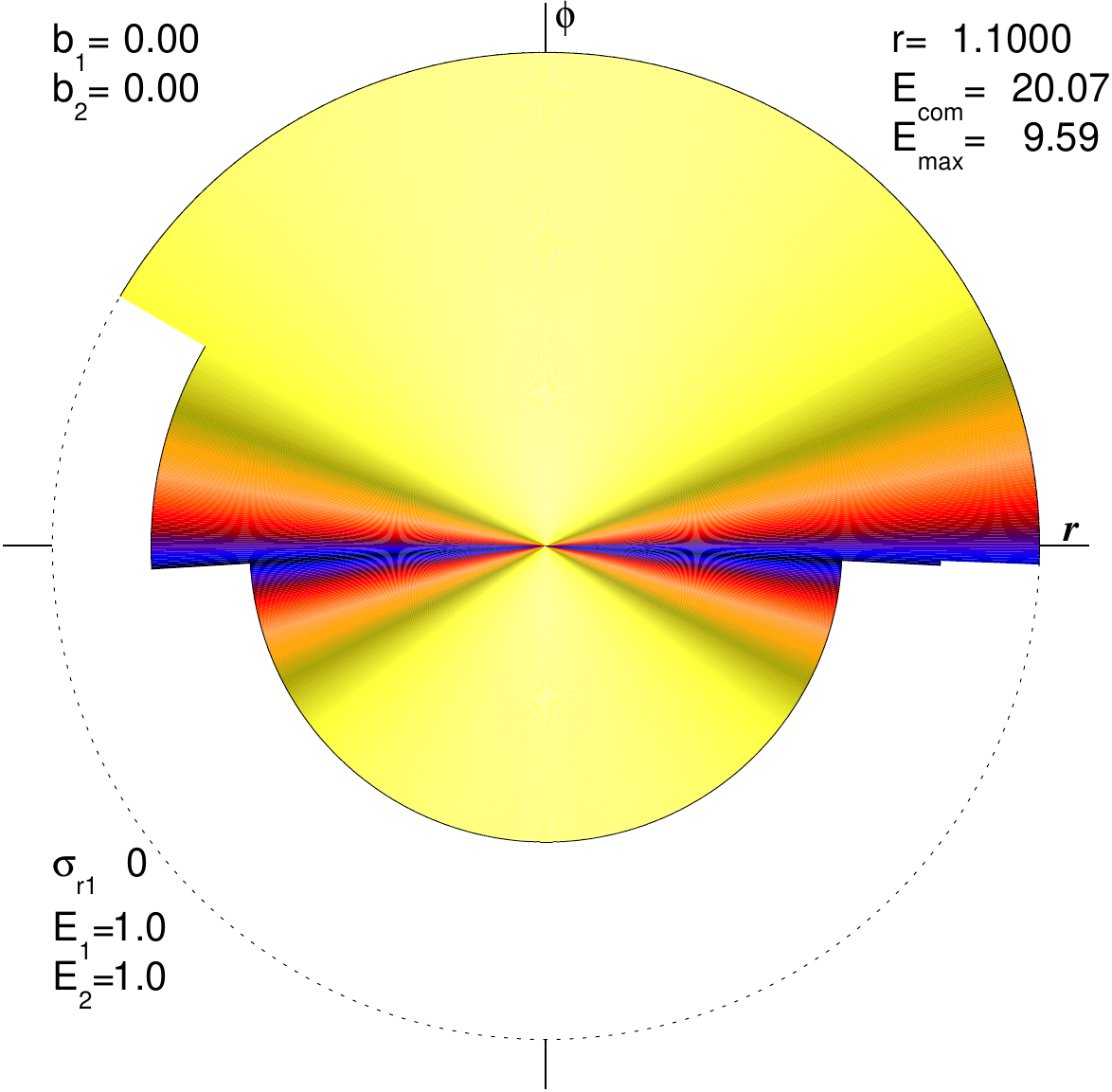}\hspace{2mm}
  \includegraphics[width=0.28\textwidth]{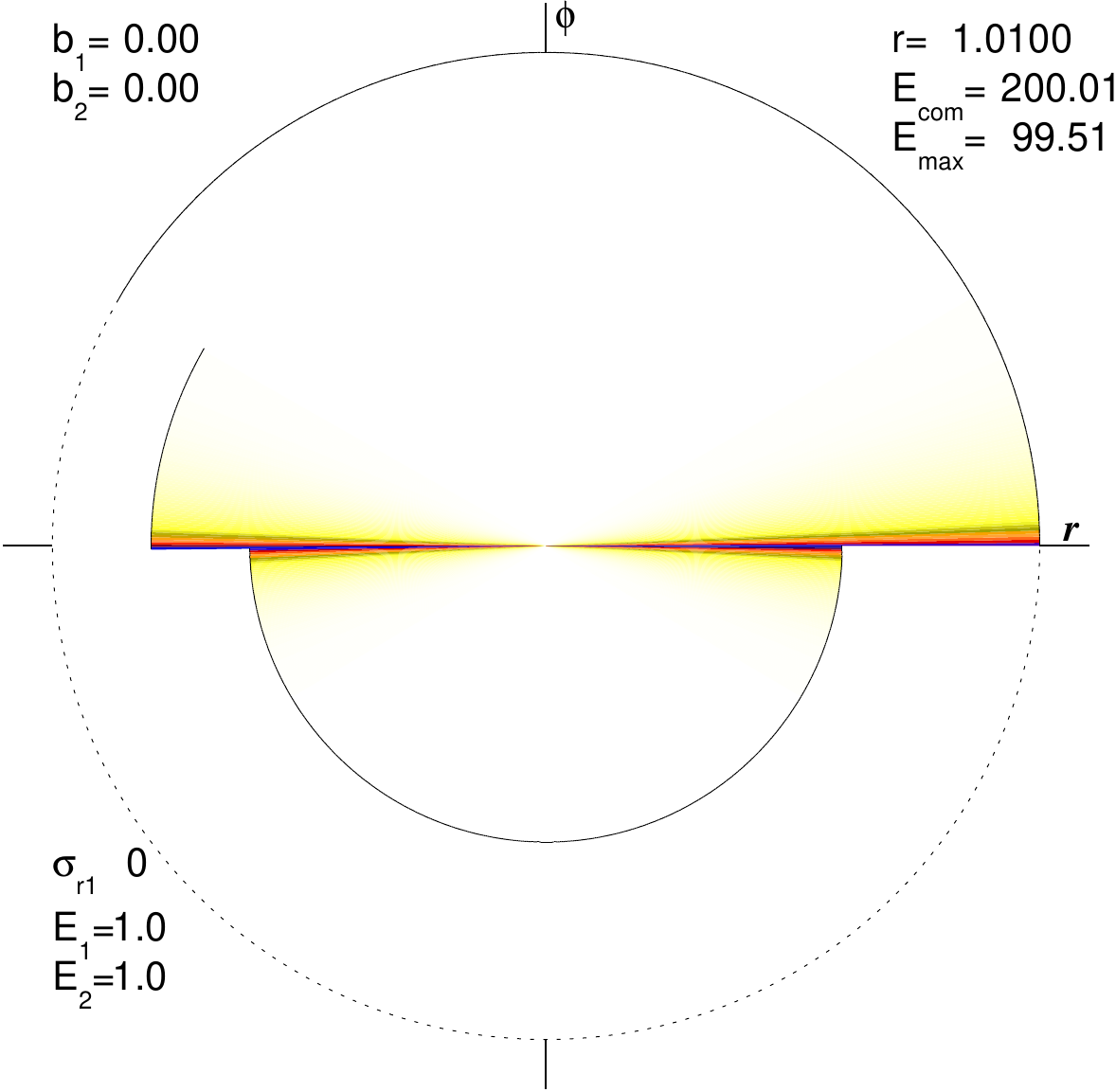}
\caption{\label{fig:polar_Berti}
  Polar plots of the energy and escape distribution of photons
  resulting from annihilations outside an extremal Kerr black hole.}
\end{center}
\end{figure}

We reproduce some of the results of \cite{Berti:2015} in Figure
\ref{fig:polar_Berti} for hypothetical massive particles with $b_1=b_2=0$
annihilating in the ergosphere, with particle 1 on an out-going
trajectory. With this selection of {\it deus ex machina} particles, it
is easy to reach very high values for $E_{\rm com}$, $\eta_{\rm max}$,
and also the escape fraction. While it is impossible for these
out-going trajectories to originate from initially infalling
particles, \cite{Berti:2015} proposes an alternative source: the
products of earlier scattering reactions. However, for the simple
cases they explore, the infalling rest mass must be sufficiently large
to produce the appropriate out-going trajectories, so that in the end,
the net efficiency is no greater than the single rebound configuration
we originally proposed in \cite{Schnittman:2014}

Before we move on to the next section, covering more generic numerical
calculations of the Penrose process, it is valuable to discuss in more
detail the analytic results of Leiderschneider \& Piran
\cite{Leiderschneider:2016}. Unlike the vast majority of papers cited
thus far, \cite{Leiderschneider:2016} extends their analysis to also
include non-planar trajectories. The reactions still take place in the
equatorial plane, but the reactant particles themselves are allowed to move out
of the plane (for the case of reactions outside of the plane, see
\cite{Gariel:2014}). In doing so, they were able to dispell one of the
popular assumptions made in many previous works: due to symmetry, the
highest energies must come from purely planar trajectories. For
example, the ``standard'' BSW case of massive, infalling planar
particles annihilating into photons just outside the horizon gives a
peak efficiency of $130\%$ \cite{Bejger:2012}. By relaxing only the
condition on the location of the collision, slightly larger values of
$b_1$ are allowed, and the resulting efficiency increases
significantly: $\eta_{\rm max} \approx 2.63$
\cite{Leiderschneider:2016} (see also \cite{Harada:2016} who correctly
identify the important problem of taking the $r,b$ limits in the
proper order, but appear to make an arithmetic error and obtain a
slightly smaller efficiency). This actually makes perfect sense: by
selecting the largest possible value for $b_1$ for a given radius, we
are in effect setting the radial velocity to zero, because that radius
corresponds to a turning point for that impact parameter. Therefore
the efficiency naturally lies somewhere between the ingoing and
outgoing results. 

Relaxing the initial conditions further, so that the incoming
particles have some motion in the $\theta$-direction, gives a larger
available center-of-mass energy, again increasing the
efficiency. However, it appears that this approach only works for
ingoing particles with $b_1=2$ and $p_r^{(1)} < 0$ (these non-planar
orbits with critical values of $b$ were also identified in
\cite{Harada:2011b}). For $p_r^{(1)}
>0$, the maximum energy and efficiency are still realized with fully
planar trajectories \cite{Leiderschneider:2016}. 

\begin{table}
\caption{\label{tab:Leider}
  Summary of results from Ref.\ \cite{Leiderschneider:2016},
  the most complete and exact work to date on maximizing energy of
  escaping particles ($E_{\rm max}$) and
  efficiency ($\eta_{\rm max}$) for the collisional Penrose process. The labeling
  convention for the particles is XYZsgn, with X, Y, and Z describing
  the properties of particles 1, 2, and 3, respectively, and 'sgn'
  describing the direction of the radial velocity for particle 1. 'M'
  refers to a massive particle falling from rest at infinity, 'P' a
  photon, and 'm' ('p') a massive particle (photon) with
  infinitessimal mass (energy) compared
  to its companion particle's energy. In all cases the first particle
  has the critical impact parameter $b_1=2$. }
\begin{tabular}{lllrlr}
\hline\noalign{\smallskip}
 & $E_{\rm max}$ & $\eta_{\rm max}$ & & $M_{3,{\rm max}}$ &  \\
\noalign{\smallskip}\hline\noalign{\smallskip}
MMP- & $2(2+\sqrt{3})$ & $2+\sqrt{3}$ & $\approx 3.73$ &  & \\
MMP+ & $(2+\sqrt{3})(2+\sqrt{2})$ & $(2+\sqrt{3})(2+\sqrt{2})/2$ & $6.37$ & & \\
PmP- & $2(2+\sqrt{3})E_1$ & $2(2+\sqrt{3})$ & $7.46$ & & \\
PmP+ & $(2+\sqrt{3})^2E_1$ & $(2+\sqrt{3})^2$ & $13.92$ & & \\
MpP- & $2(2+\sqrt{3})$ & $2(2+\sqrt{3})$ & $7.46$ & & \\
MpP+ & $(2+\sqrt{3})(2+\sqrt{2})$ & $(2+\sqrt{3})(2+\sqrt{2})$ & $12.74$ & & \\
\noalign{\smallskip}\hline\noalign{\smallskip}
MMM- & $4+\sqrt{11}$ & $(4+\sqrt{11})/2$ & $3.66$ & $2\sqrt{3}$ &
$\approx 3.46$\\
MMM+ & $7+4\sqrt{2}$ & $(7+4\sqrt{2})/2$ & $6.32$ & $\sqrt{3}(2+\sqrt{2})$ & $5.91$\\
PmM- & $4E_1+\sqrt{(12E_1^2-1)}$ & $2(2+\sqrt{3})$ & $7.46$ &
$2\sqrt{3}E_1$ & $5.91E_1$\\
PmM+ & $2(2+\sqrt{3})E_1+$        & $(2+\sqrt{3})^2$ & $13.92$ &
$\sqrt{3}(2+\sqrt{3})E_1$ & $6.46E_1$ \\
     & $\sqrt{3(2+\sqrt{3})^2E_1^2-1}$ & & & & \\
MpM- & $4+\sqrt{11}$ & $4+\sqrt{11}$ & $7.32$ & $2\sqrt{3}$ & $3.46$\\
MpM+ & $7+4\sqrt{2}$ & $7+4\sqrt{2}$ & $12.66$ &
$\sqrt{3}(2+\sqrt{2})$ & $5.91$\\

\noalign{\smallskip}\hline
\end{tabular}
\end{table}

In Table \ref{tab:Leider} we reproduce a summary of the analytic
results for peak efficiency from Ref.\ \cite{Leiderschneider:2016},
combining their Tables
1 and 2. We follow their notation describing the parameters of
collisions as massive particles 'M' and 'm', photons 'P' and 'p', and the
direction of particle 1 (positive or negative radial
velocity). When particle 2 has mass 'm', this means that one should
take the limit of $E_1 >> m$. Similarly, when particle 2 is a photon
of energy 'p', this corresponds to $M_1 >> p$ (these results for
``heavy'' massive particles were derived independently by
\cite{Zaslavskii:2016b}). 
For massive products, we also list the peak rest mass
attainable for particle 3, which does not necessarily correspond to
the same trajectories used to achieve peak efficiency
\cite{Leiderschneider:2016}. 

As can be seen from the results in Table \ref{tab:Leider}, the
absolute maximum efficiency is still that discovered in
\cite{Schnittman:2014}. But we also see that generally,
high-efficiency collisions can be realized for a wide variety of
generic reactions. The unifying theme appears to be the critical
angular momentum for particle 1, along with near-horizon collisions
around extremal black holes. 

\section{Numerical Calculations}\label{sec:numeric}
Compared to the extensive analytic work described in the previous
section, there have been significantly fewer numerical studies of the
Penrose process. Yet for just about any astrophysical application,
full 3D (6D in phase space) calculations are required to predict
observable features of the reaction products. Astrophysical
applications will also require the use of generic spin parameters,
as opposed to the maximal spin limit. 

\begin{figure}
\begin{center}
  \includegraphics[width=0.45\textwidth]{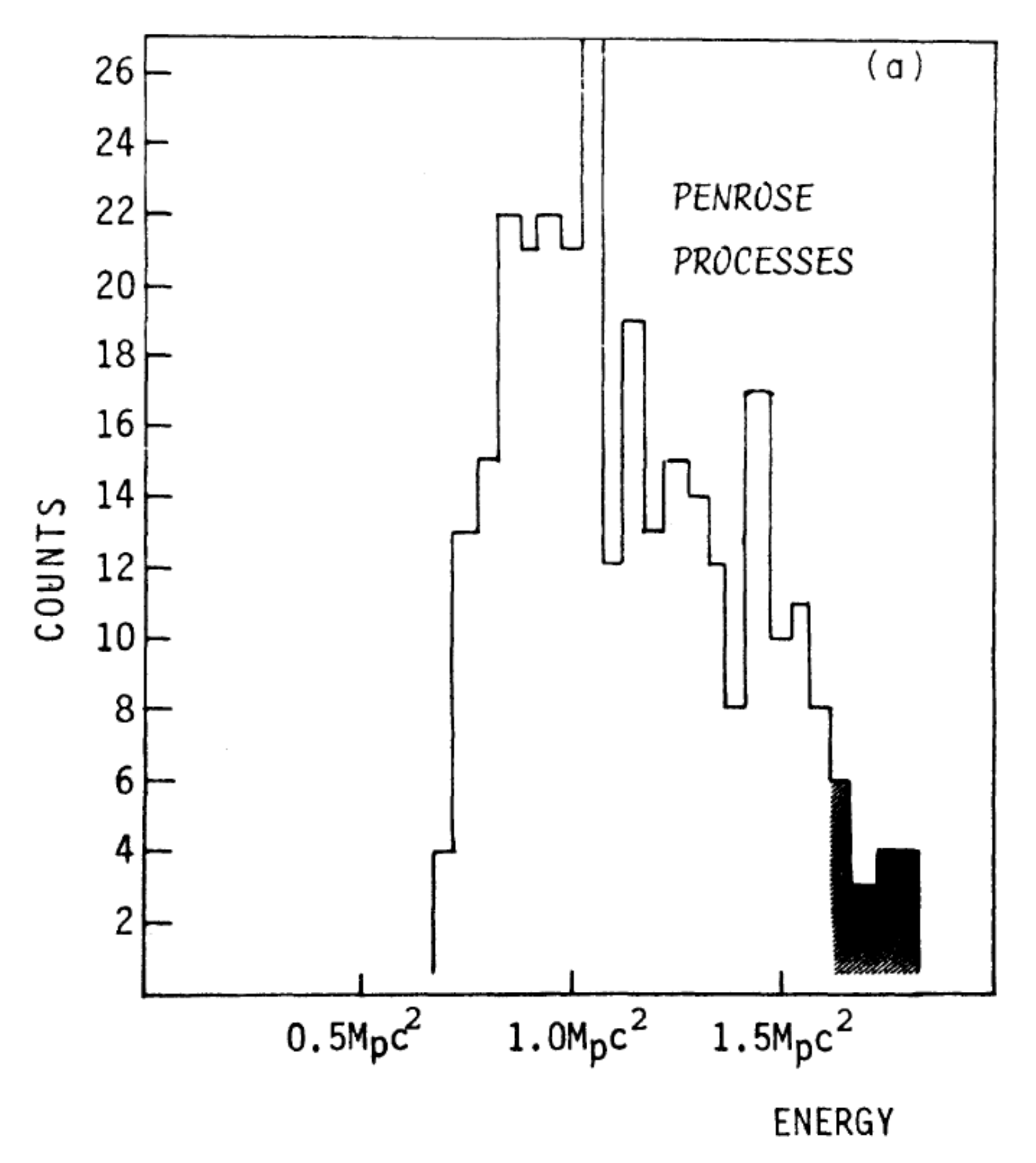}
\caption{\label{fig:Piran77_5a}
  Spectrum of outgoing massive particles from a numerical
  scattering experiment including 1000 elastic collisions between
  protons with particle 1 falling in radially from infinity and
  particle 2 on a circular orbit at the ISCO for a black hole with
  spin $a_\ast=0.998$. Of the 2000 protons taking place in the
  scattering events, only 23 escape with $E_3>E_1+E_2$. Reproduced
  from \cite{Piran:1977}.}
\end{center}
\end{figure}

The first such attempt
at a numerical calculation was done in the seminal paper by Piran and
Shaham \cite{Piran:1977}, where they did a Monte Carlo simulation of
the particles produced by elastic scattering of infalling
protons with identical particles on stable circular planar
orbits. While extremely 
impressive for the time, computational limitations restricted their
calculations to a few thousand protons, enough to get a qualitative
feel for the spectral properties and Penrose process rates, but hardly
enough to fully sample the phase space. One representative example is
shown in Figure \ref{fig:Piran77_5a}. This shows the outgoing energy
spectrum for particle 3, also massive in this example. The black hole
spin is $a_\ast=0.998$, particle 2 is on a
bound circular orbit at the ISCO, and particle 1 falls in from
infinity with zero angular momentum. Of the 2000 protons participating
in the 1000 collisions included in their calculation, only 23 escape
with energy greater than $E_1+E_2=1.674M_pc^2$. 

Nearly 20 years later, with significant advances in computing power, a
more comprehensive study was carried out in \cite{Williams:1995},
covering a wider range of collisional cases, including pair production,
Compton scattering, and gamma-ray-proton pair production 
($\gamma+p \to p+e^-+e^+$). Again, the focus was on astrophysical
applications, so the canonical spin of $a_\ast=0.998$ was used. This
work was further expanded in \cite{Williams:2002}, exploring the range
of angles and energy for outgoing photons. 

While these earlier works were able to explore a much wider range of
parameter space for the collision products, they were still generally
limited to a relatively small number of specific {\it initial} conditions
for the reactants. In \cite{Schnittman:2015} this author attempted to
expand on this approach and carry out a numerical calculation of the
full 6D distribution of both reactants and products for annihilation
events around spinning black holes. Using the radiation transport code
\pan\, to integrate geodesic trajectories, we populate the phase space
by launching a large number of test particles around the black hole.
At each time step along the trajectory, a weighted contribution is
added to the 6-dimensional phase space (in practice, ``only''
5-dimensional, because of the azimuthal spatial symmetry of the Kerr
metric).  

We divide the distribution
into two populations: bound and unbound. The unbound population has a
thermal, non-relativistic velocity distribution at infinity. The bound
population is constructed to produce a power-law slope in density, and
only includes particles on stable orbits with specific energy less
than unity, isotropic as seen local quasi-stationary observer, with a
Maxwell-Juttner velocity distribution and characteristic energy
corresponding to a virial temperature. In principal, any slope can be
produced, but we generally restrict ourselves to $\rho \sim r^{-2}$
following \cite{Gondolo:1999,Sadeghian:2013}.  

\begin{figure}
\begin{center}
  \includegraphics[width=0.45\textwidth]{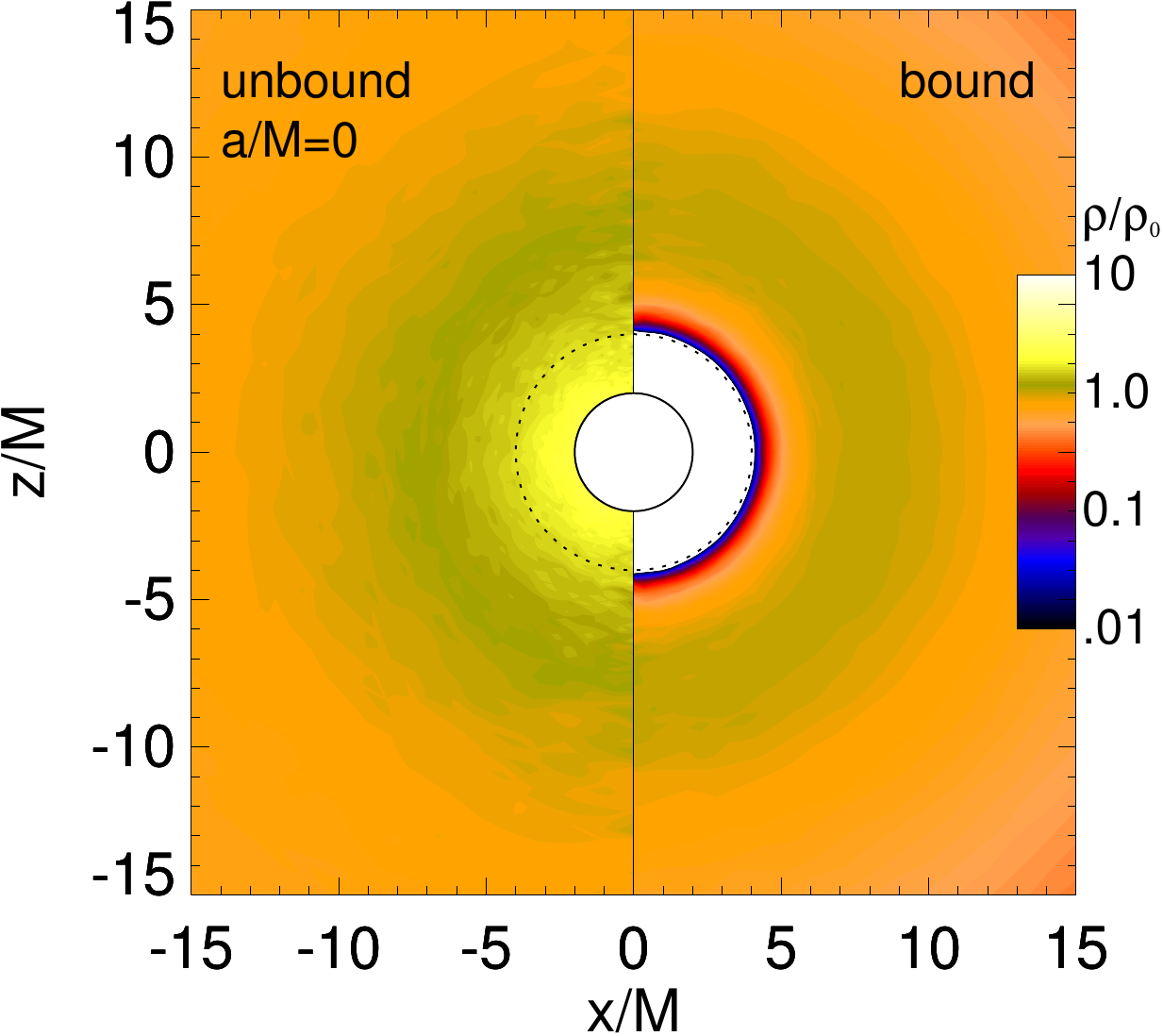}
  \includegraphics[width=0.45\textwidth]{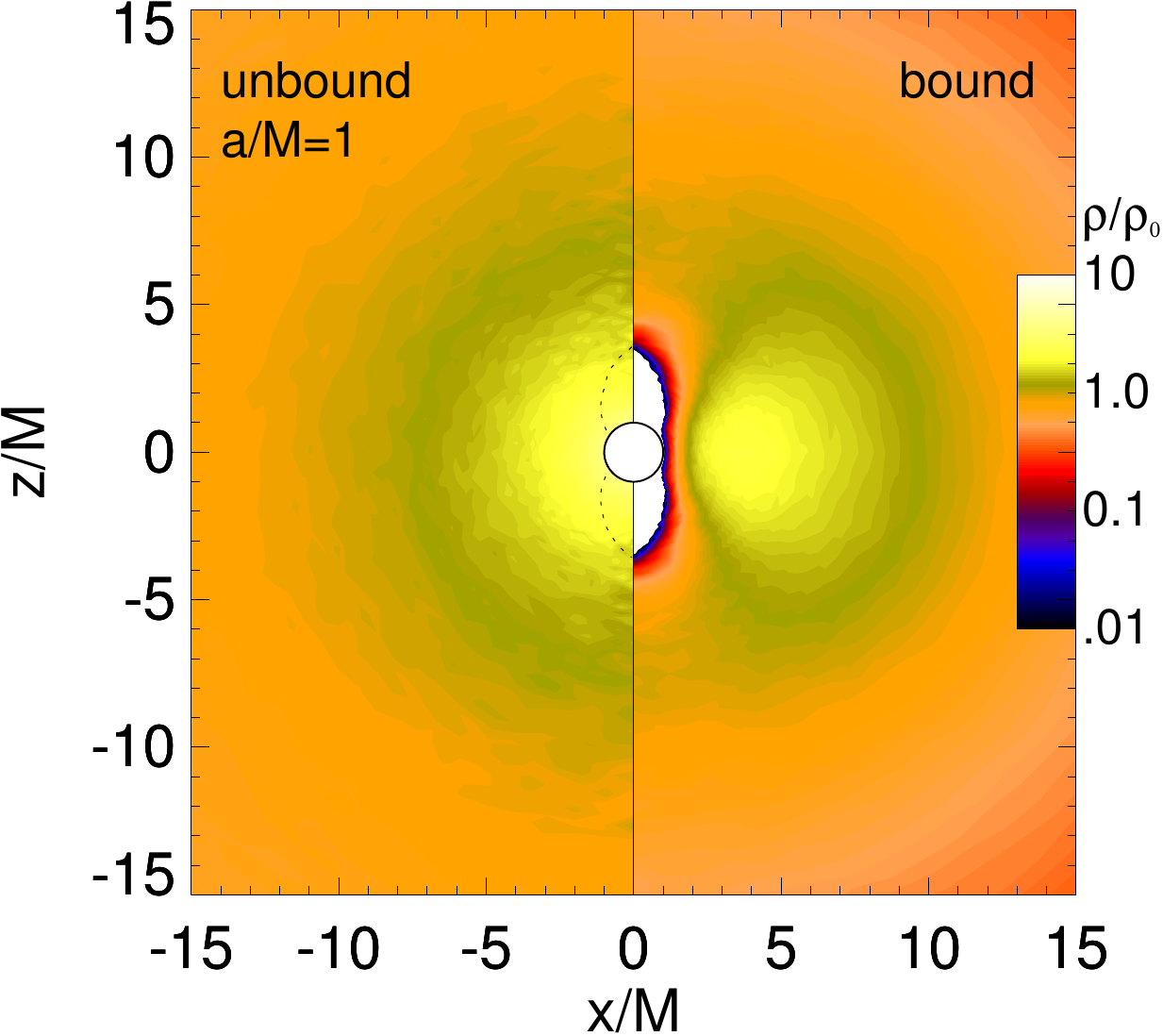}
\caption{\label{fig:JS_f8}
  Spatial density of test particles in
  the $x-z$ plane, for both bound and unbound populations, for $a/M=0$
  and $a/M=1$. For each case, we show the
  unbound distribution on the left side and the bound distribution on
  the right side of the plot, and all distribution functions are
  normalized to the mean density at $r=10M$.
  The horizon is plotted as a solid curve
  and the radius of the marginally bound orbits is shown as a dotted
  curve. The spin axis of the black hole is in the $+z$
  direction. Reproduced from \cite{Schnittman:2015}.}
\end{center}
\end{figure}

The main results of this calculation are shown in Figure
\ref{fig:JS_f8}, reproduced from \cite{Schnittman:2015}. The contour
plots show 2D cuts in the $(r,\theta)$ plane of the density
distribution for bound and unbound populations, for spin parameters of
$a_\ast=0$ and $a_\ast=1$. Because of the numerical nature of this
approach, any spin can be used, but these obviously span the range of
astronomical possibilities. The density distribution of the bound
population agrees closely with the analytic results of
\cite{Sadeghian:2013} for non-spinning black holes, and
\cite{Ferrer:2017} for the Kerr case. It is interesting to note that
for the spinning case, the unbound density distribution is almost
perfectly uniform in $\theta$, rising steadily towards the horizon, despite the
fact that many of these particles spend a large amount of coordinate
time orbiting near the midplane before finally plunging. On the other
hand, the bound population shows a clear break in symmetry, due to the
increased stability of prograde, planar orbits. These orbits
contribute to a density spike in the form of a thick torus, peaking
around radius $r=4M$ \cite{Schnittman:2015}.

\begin{figure}[h]
\caption{\label{fig:JS_f5} Momentum distribution of unbound
  particles observed by a FFIO in the equatorial plane at radius
  $r=2M$. All particles have nearly unitary specific energy at
  infinity, so the average particle speed is on the order
  $\sqrt{2GM/r} \approx c$ (panel {\it a}). In panels ({\it
  b-d}) we show the distribution of the individual momentum
  components, which are decidedly non-thermal and highly
  anisotropic. Reproduced from \cite{Schnittman:2015}.}
\begin{center}
  \includegraphics[width=0.4\textwidth]{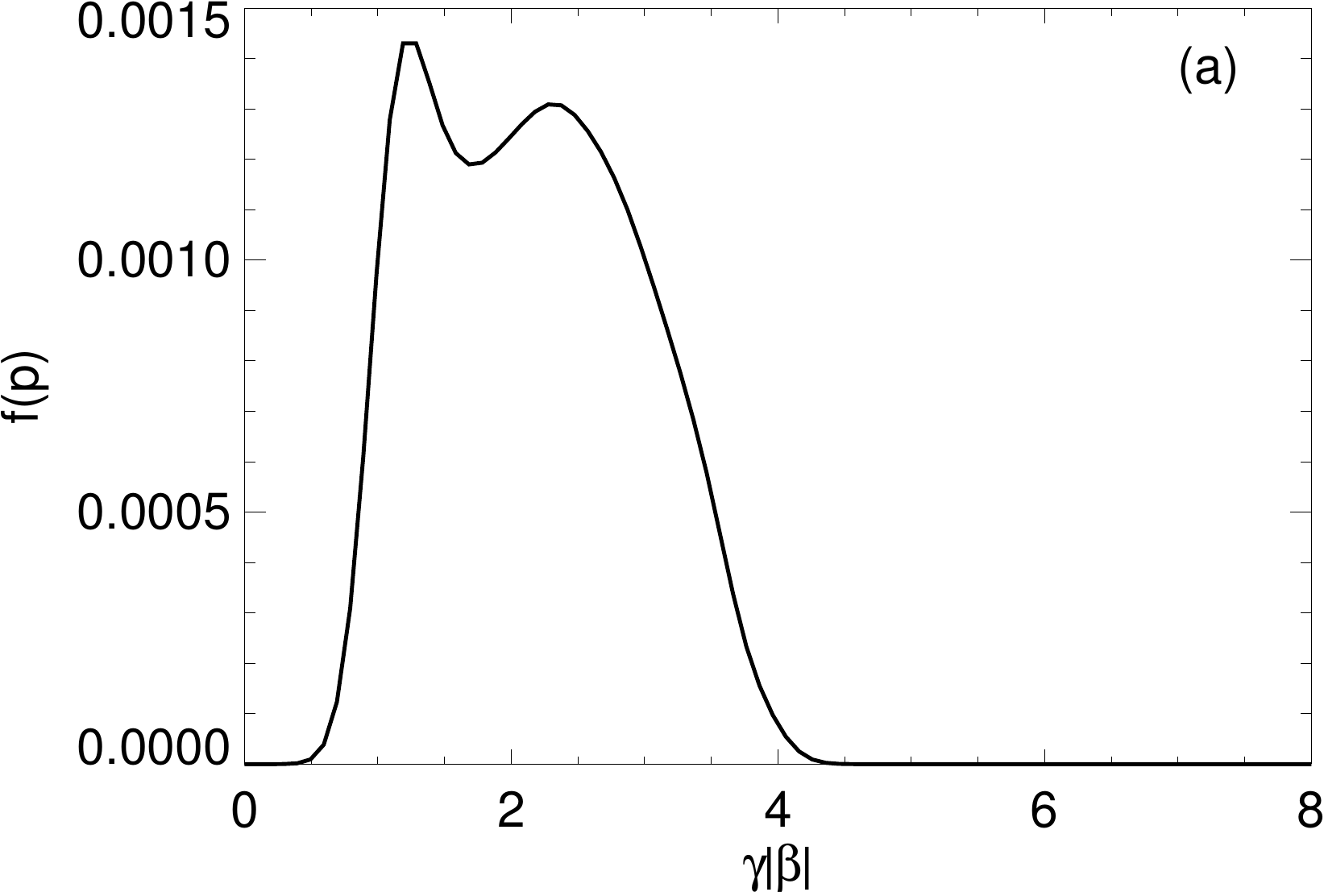}
  \includegraphics[width=0.4\textwidth]{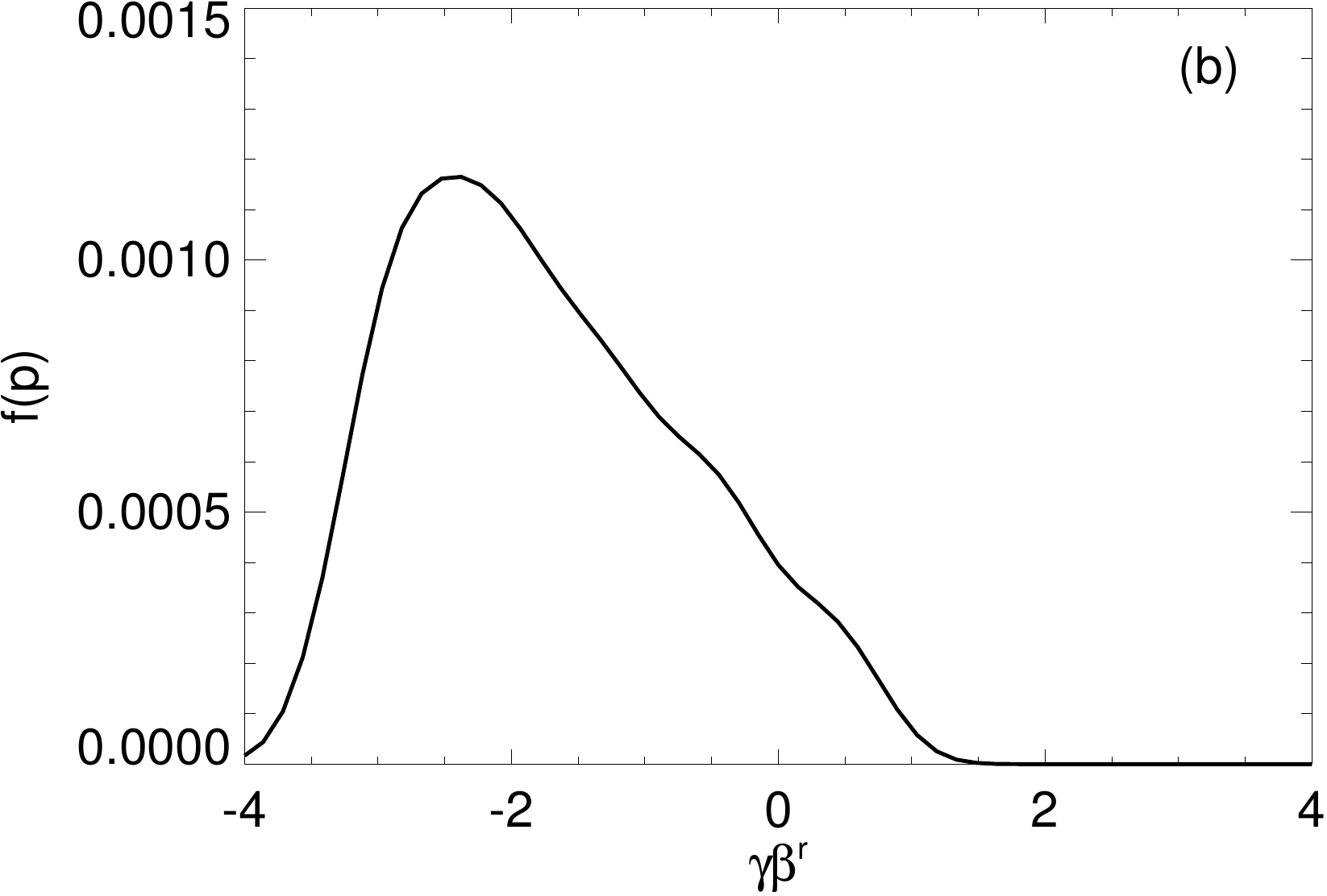} \\
  \includegraphics[width=0.4\textwidth]{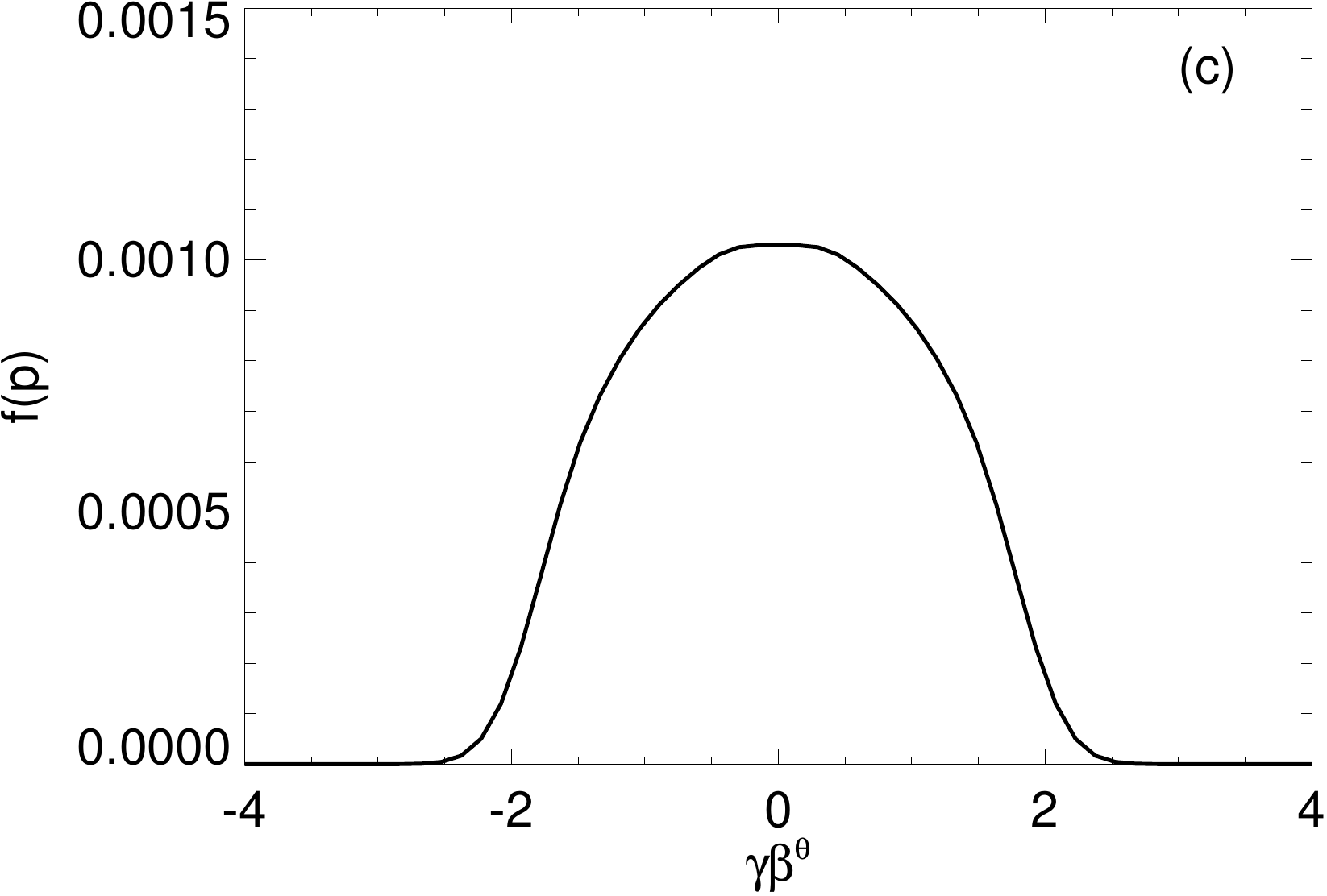}
  \includegraphics[width=0.4\textwidth]{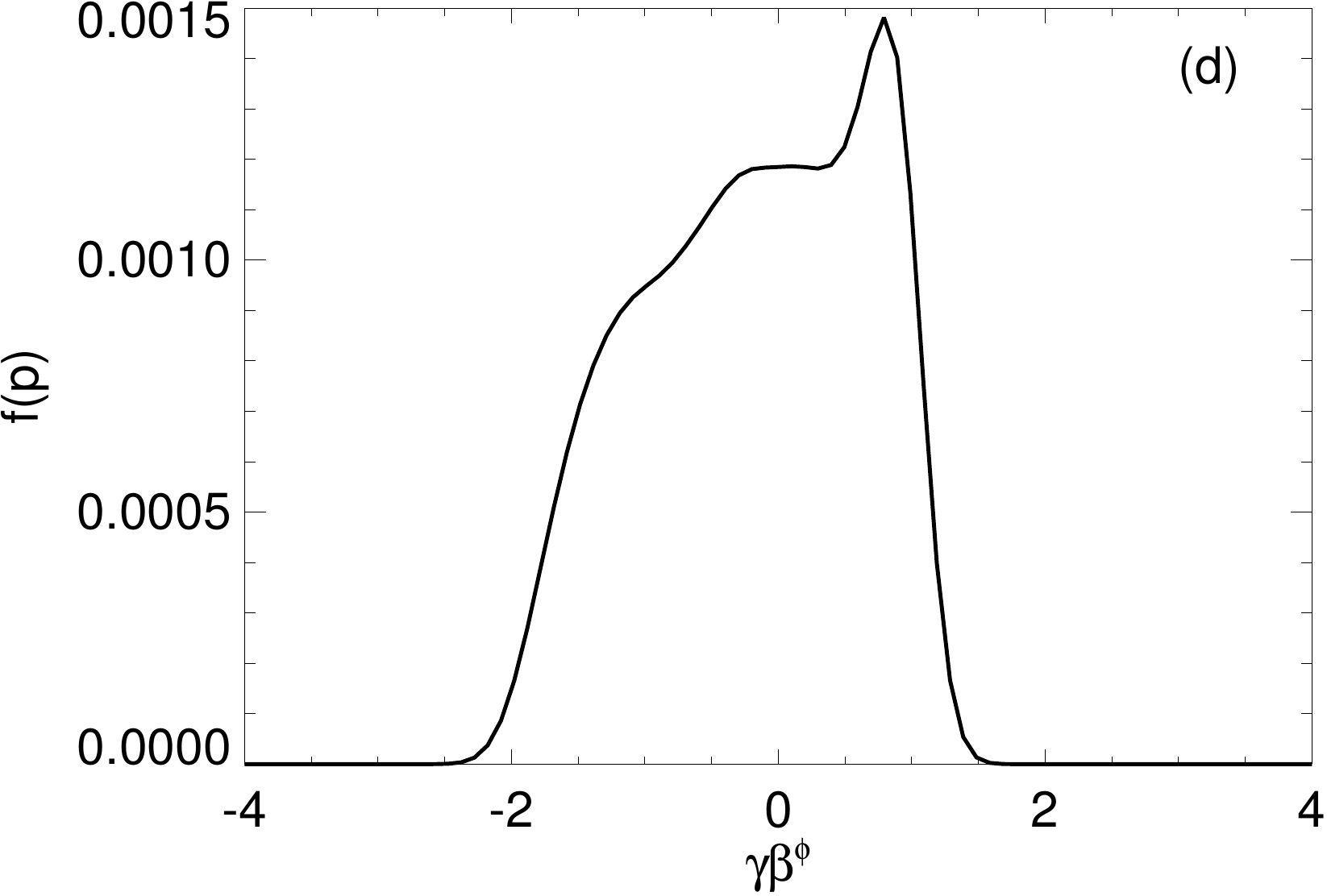}
\end{center}
\end{figure}

\begin{figure}[h]
\caption{\label{fig:JS_f10} Momentum distribution of bound
  particles measured by a LNRF observer in the equatorial plane at radius
  $r=2M$. Compared to Figure \ref{fig:JS_f5}, here we actually
  see a {\it more}
  symmetric, thermal distribution making up a thick torus of stable,
  roughly circular orbits near the equatorial plane. Reproduced from
  \cite{Schnittman:2015}.}
\begin{center}
  \includegraphics[width=0.4\textwidth]{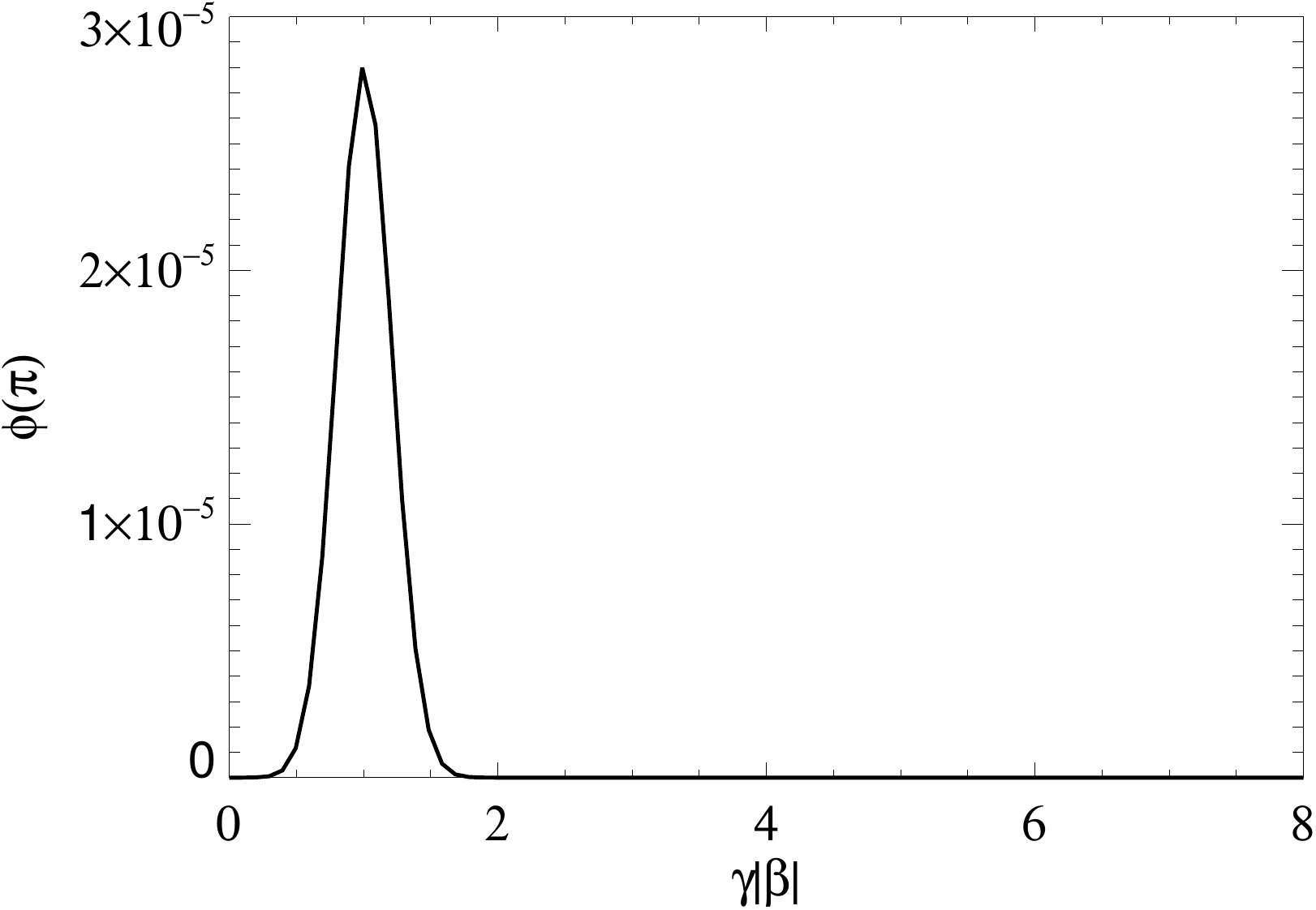}
  \includegraphics[width=0.4\textwidth]{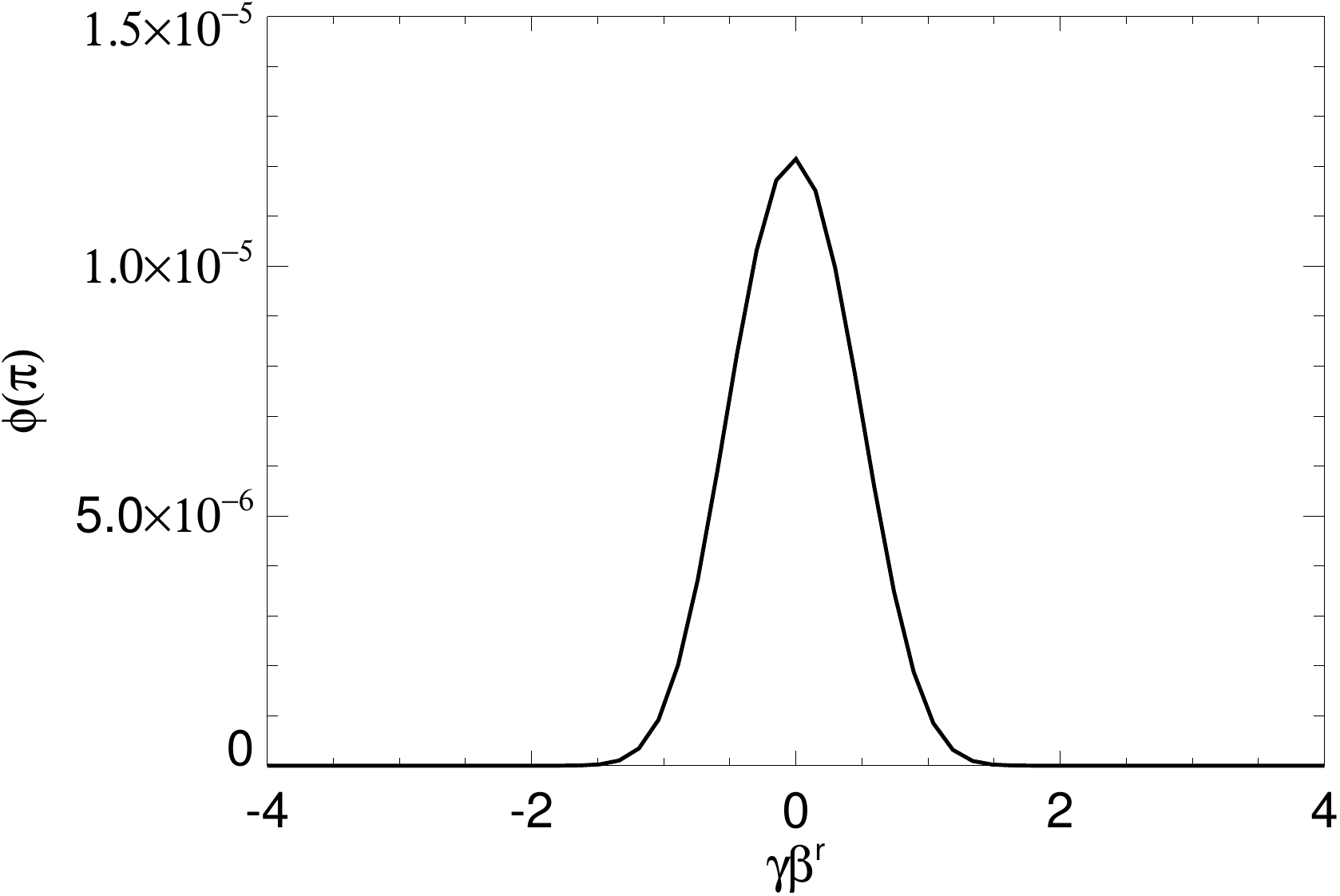} \\
  \includegraphics[width=0.4\textwidth]{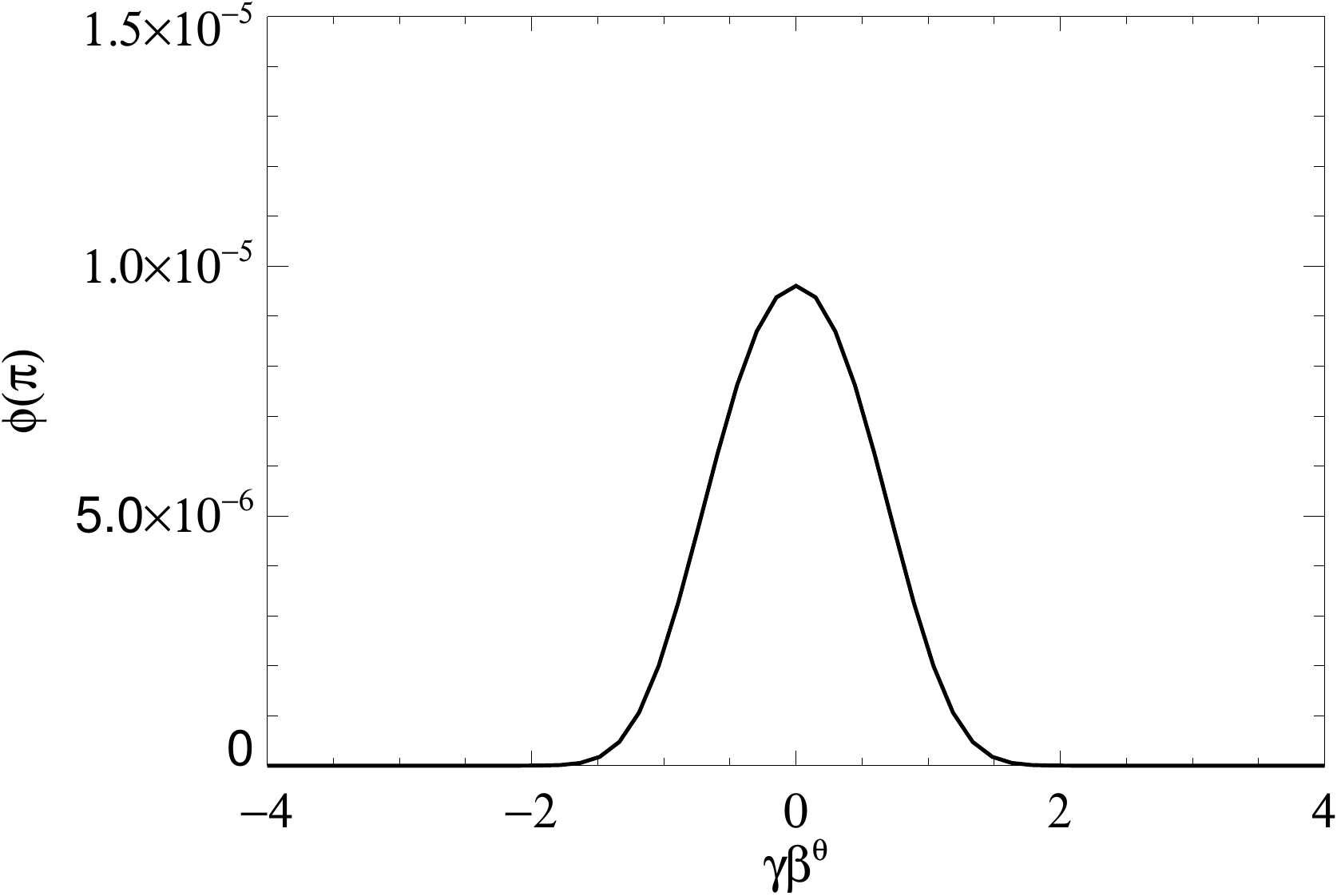}
  \includegraphics[width=0.4\textwidth]{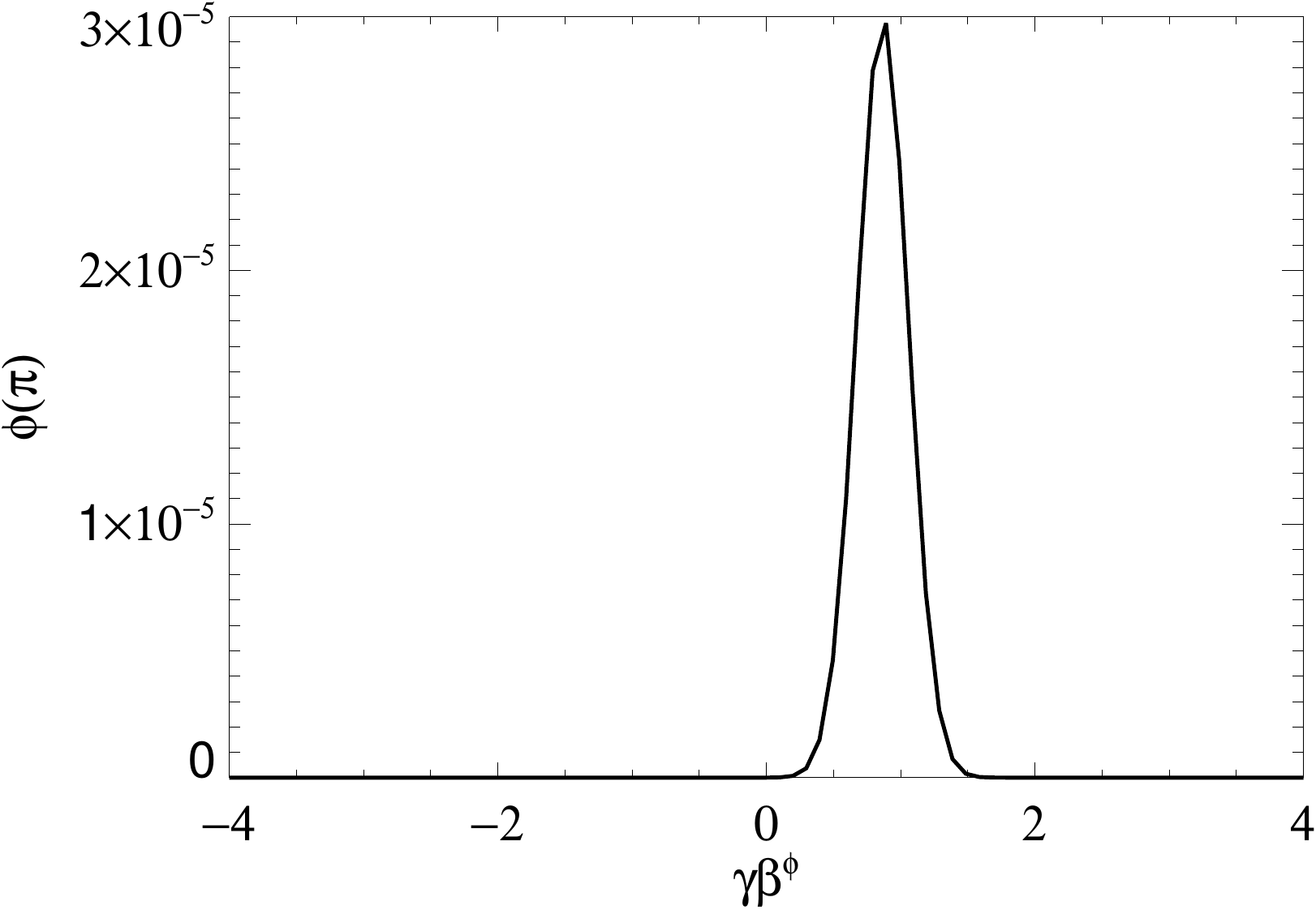}
\end{center}
\end{figure}

In addition to the density distribution $\rho(r,\theta)$, we also
calculate the distribution in velocity space at each point. These
results are shown in Figures \ref{fig:JS_f5} and \ref{fig:JS_f10} for
the unbound and bound populations, respectively. In all cases, the
velocities are measured by an observer in the equatorial plane at
radius $2M$. For the unbound population, the observer is
free-falling from infinity (FFIO) with zero angular momentum; for the bound
population, the observer has no radial motion, but rotates with the
spin of the black hole despite zero angular momentum (LNRF, locally
non-rotating frame in the language of \cite{Bardeen:1972}). 

Despite the fact that the spatial density distribution for unbound
particles appears quite uniform in $\theta$ in Figure \ref{fig:JS_f8},
we see that the velocity distribution near the black hole is not at
all isotropic. There is essentially a bimodal distribution of
velocity, with retrograde particles plunging with large negative
values of $v^r$, and prograde particles corating with the black hole
spin, peaked around $v^r=0$ and $\gamma v^\phi=c$. 

As can be seen in Figure \ref{fig:JS_f10}, the bound population is
much more isotropic. This is hardly surprising, as there are
no stable retrograde orbits at $r=2M$, and even the prograde orbits
are almost perfectly planar and circular, spanning a narrow range of
velocity as seen by a nearly stationary, LNRF observer. 

\section{Dark Matter Applications}\label{sec:DM}

Despite the wide variety of fundamental and fascinating results
described in the previous sections, by most accounts the
collisional Penrose process is unlikely to play a significant role in
astrophysical processes. Even the highest efficiency reactions can
only provide energy boosts on the order of a factor of ten or so, far
below the ultrarelativistic particles seen from gamma-ray bursts or active
galactic nuclei\footnote{A leading theory for magnetically powered
  jets is the Blandford-Znajek process \cite{Blandford:1977}, which
  does extract energy from
  the spin of the black hole, but not through a particle-based Penrose
  process.}. And in any case, even those moderately high-efficiency
events require 
such fine tuning of initial conditions, they are probably impossible
to realize in a natural setting. 

One potential (although admittedly speculative) exception is the
annihilation of dark matter (DM) particles in the ergosphere around a
Kerr black hole. Numerous authors have pointed out the important role
that supermassive black holes might play in shaping the DM density
profile around galactic nuclei
\cite{Gondolo:1999,Merritt:2002,Fields:2014,Shapiro:2014,Shapiro:2016}. However,
in almost all these cases, the enhanced DM density---and thus
annihilation signal---is a purely Newtonian effect, and is therefore
not within the scope of this review. 

However, if the DM density profile is sufficiently steep, or the
annihilation cross section increases with energy (e.g., through p-wave
annihilation \cite{Feng:2010}), then the
annihilation signal will be dominated by reactions closest to the
black hole, where relativistic effects {\it are} important. With the
full phase-space distribution function calculated as in the previous
section, in \cite{Schnittman:2015} we were able to calculate the
outgoing spectrum from a sample of possible annihilation models. One
simple model is where the cross section for annihilation increases
greatly above a certain threshold energy, analogous to pion creation
via proton-proton reactions. With the Monte Carlo code \pan, we can
sample the phase space of test particles from both bound and unbound
populations, and calculate annihilation rates for a given cross
section model. The simplest annihilation model produces two photons of
equal energy and isotropic in angle in the center-of-mass frame of the
reacting DM particles. These photons are then propogated to an
observer at infinity, where they can be summed to produce images and
spectra. 

\begin{figure}[ht]
\caption{\label{fig:JS_f13} Simulated image of the annihilation
  signal around an extremal Kerr black hole, considering only
  annihilations from unbound DM particles with $E_{\rm com} > 3m_\chi$. The
  observer is located in the equatorial plane with the spin axis
  pointing up. While the image appears off-centered, it is actually
  aligned with the coordinate origin. Reproduced from \cite{Schnittman:2015}.}
\begin{center}
  \includegraphics[width=0.7\textwidth]{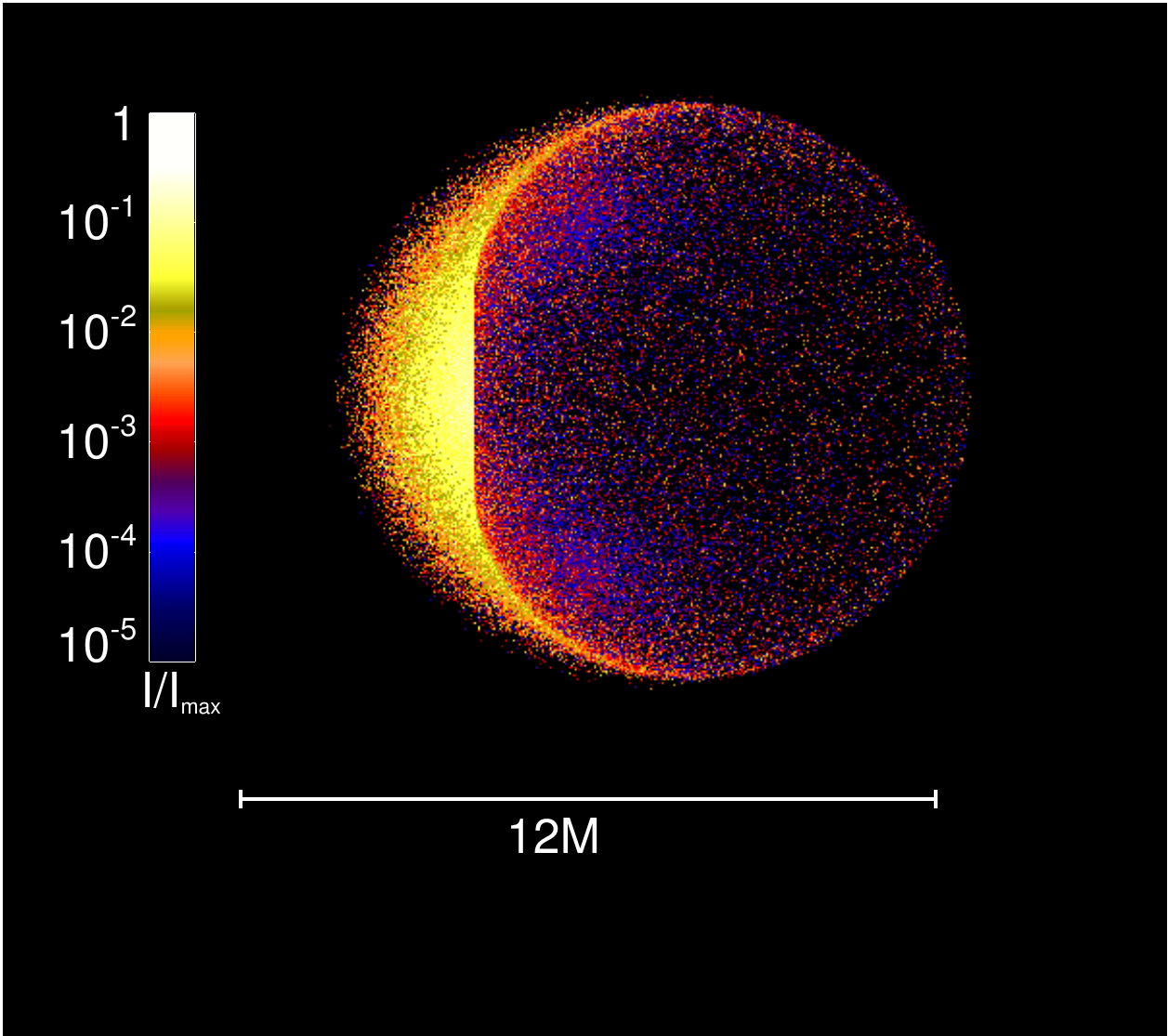}
\end{center}
\end{figure}

If we take the threshold center-of-mass energy to be a moderate $3
m_\chi c^2$, we find that most of the annihilation photons are
produced within the ergosphere region, and are thus sensitive probes
of the Penrose process. In Figure \ref{fig:JS_f13} we show a simulated
image produced by the annihilation photons produced from the unbound
DM population around an extremal Kerr black hole, as seen by an
observer at infinity and inclination of $90^\circ$ from the spin
axis. The extreme frame dragging and Doppler boosting from prograde
orbits make the image highly asymmetric. Clearly visible is the
characteristic shadow of a Kerr black hole, with a flattened prograde
edge, as described in \cite{Chandra:1992}. 

\begin{figure}[ht]
\caption{\label{fig:JS_f16} Observed flux from annihilation products
  near a black hole, as a function of spin parameter. (left)
  Contribution from the unbound population, including only
  annihilations with $E_{\rm com}>3m_\chi c^2$. (right) The bound
  population, with $\rho \sim r^{-2}$ and no threshold energy. In all
  cases the observer is in the equatorial plane. The scale of the
  vertical axis is arbitrary.
  Reproduced from \cite{Schnittman:2015}.}
\begin{center}
  \includegraphics[width=0.45\textwidth]{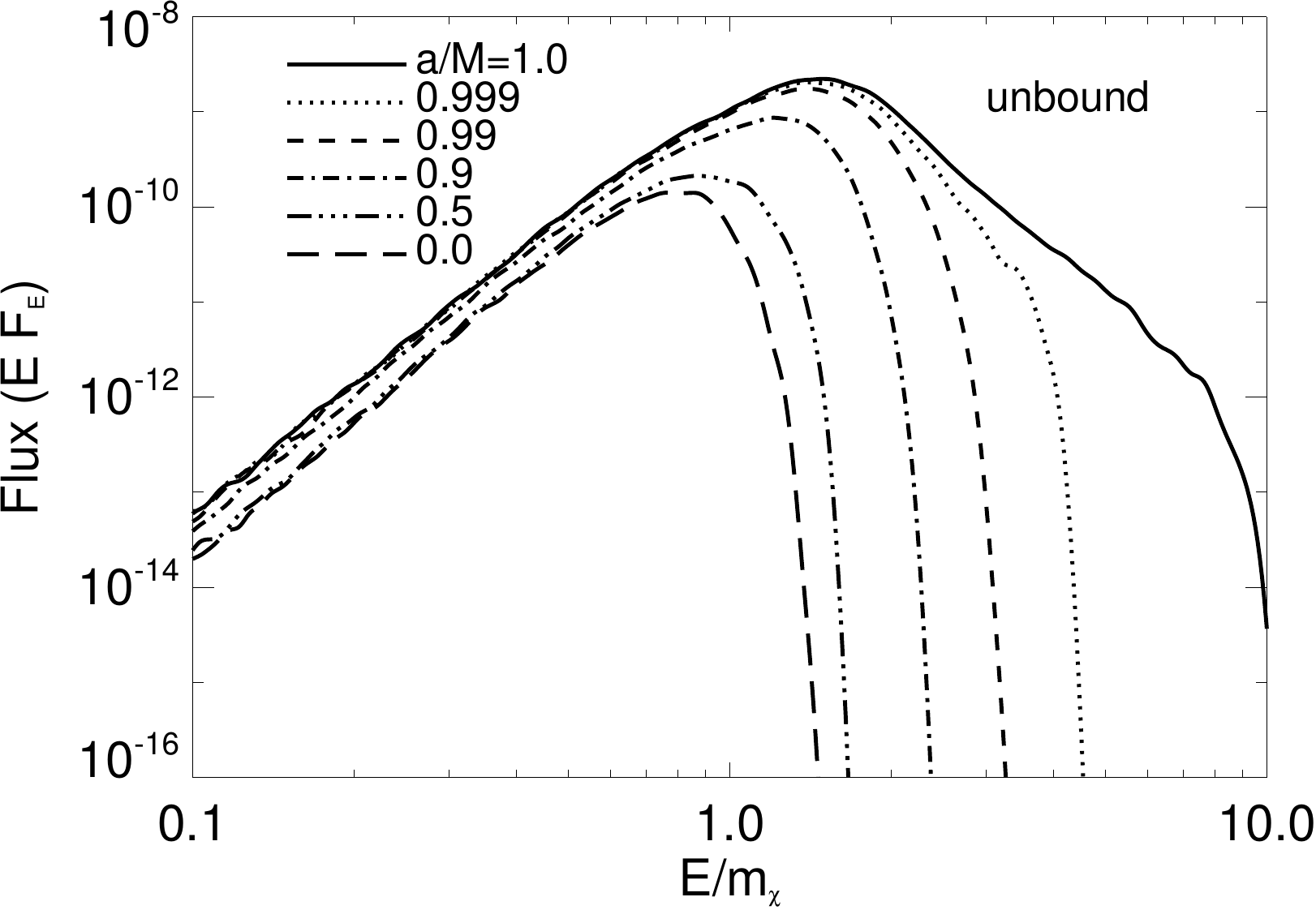}
  \includegraphics[width=0.45\textwidth]{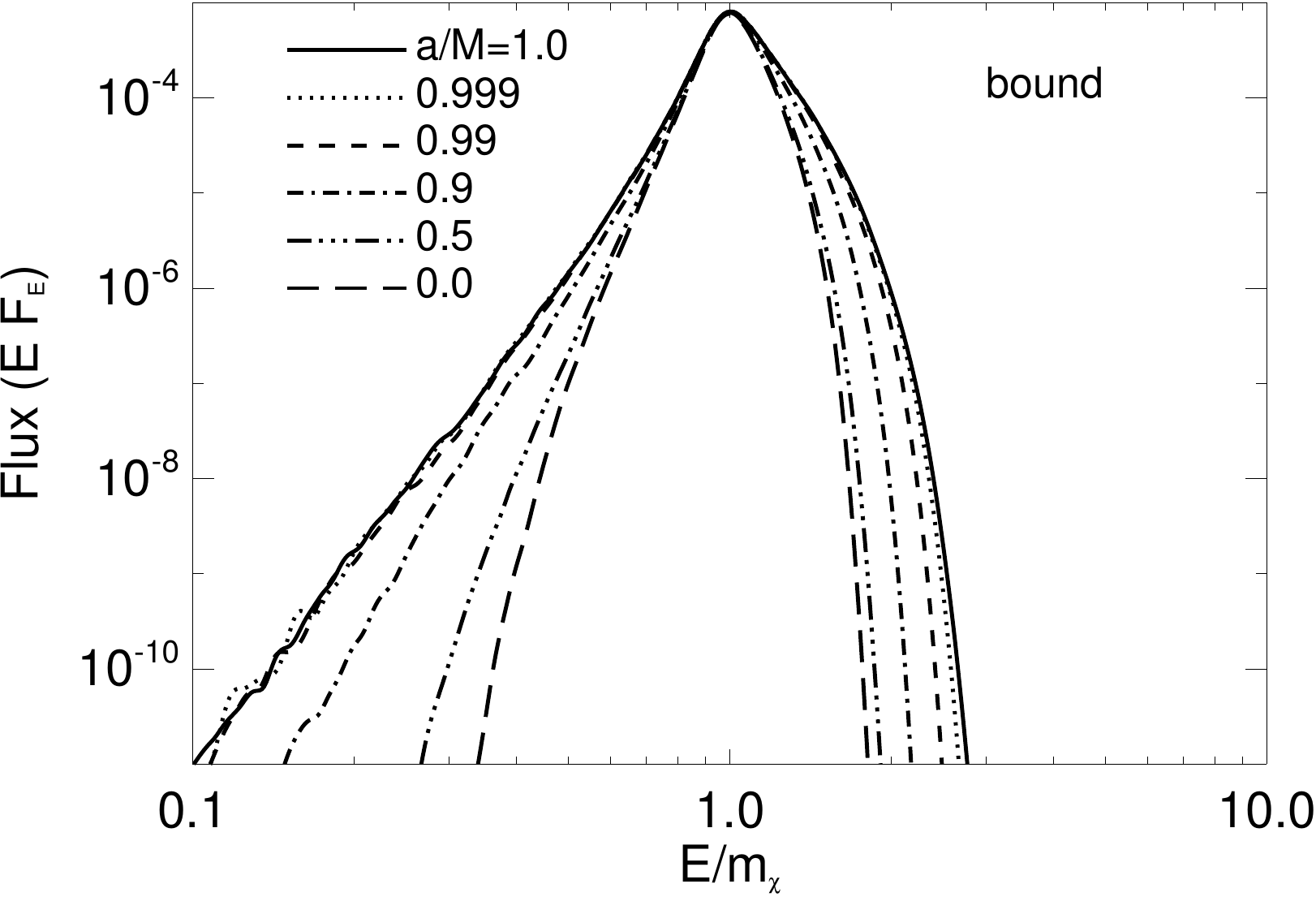}
\end{center}
\end{figure}

In Figure \ref{fig:JS_f16} we show the spectra corresponding to this
annihilation scenario, for a range of black hole spins, for both the
unbound and bound populations. In order to highlight the effects of
the black hole spin, we focus on reactions coming from close to the
black hole. For the unbound population, this means using an energy
threshold for the annihilation cross section of $E_{\rm com} > 3m_\chi
c^2$. For the bound population, no threshold is needed, as the density
peak near the black hole naturally leads to the annihilation signal
being dominated by photons coming from small $r$. Note the
qualitatively different spin dependences in the two cases: for unbound
DM particles, the low energy part of the spectrum is independent of
spin, as all these photons come from plunging particles near the
horizon, and experience significant gravitational redshift. At the
high energy end, we see the clear importance of spin in generating
high-efficiency, extreme Penrose process reactions. For the bound
population, on the other hand, the stable orbits do not intersect with
very large c.o.m. energies, so even for very high spins we do not see
much influence from the Penrose process. Yet the spin does play an
important role in shaping the low-energy end of the spectrum, as
higher spins allow stable orbits closer to the horizon, and thus more
extreme gravitational redshift, just as in the case of the red tail of
the iron flourescent lines seen around black holes of all sizes
\cite{Reynolds:2003}.  

The overall vertical axes in Figure \ref{fig:JS_f16} are arbitrary,
because we still don't know very much about the overall density
scaling of DM distributions around black holes. Even more uncertain is
the amplitude of the annihilation cross section, much less its energy
dependence. We hope that in the future, as gamma-ray telescopes
improve in angular and energy resolution, we will be able to use
quiescent black holes in galactic nuclei to probe the properties
of the DM particle, measure black hole spins, and explore the exotic
physics that describe the ergosphere. Perhaps one day we might even
discover advanced civilizations that have successfully harnessed the
black hole spin as an energy source, as imagined by Penrose
in his original paper \cite{Penrose:1969}!

\section{Discussion}\label{sec:discussion}

We have provided a broad overview of some of the recent work on the
collisional Penrose process, with particular focus on collisions
around extremal Kerr black holes. Despite the numerous astrophysical
limitations, since the publication of BSW \cite{Banados:2009}, there
has been a great deal of interest in determining the highest
attainable collision efficiencies. These high-efficiency reactions
require both large center-of-mass energies and also fine tuning of the
reaction product trajectories in order to assure they can escape from
the black hole. While non-Kerr (or even non-GR) black holes could more
generally lead to diverging center-of-mass energies, we have
restricted this review to classical, if extremal, Kerr black holes. 

For more general astrophysical observations, dark matter annihilation
appears to be one of the more promising applications of the Penrose
process. One reason for this is that DM particles are most likely to
travel along perfect geodesics, even in the presence of the diffuse gas
and strong magnetic fields typically found around astrophysical black
holes. Additionally, the DM density distribution is expected to peak
near galactic nuclei, which also contain supermassive black
holes. Lastly, the extreme gravitational field of the black hole is a
promising mechanism to enhance annihilation, both through increasing
the relative collision energy, and also through gravitational focusing
that increases the DM density. 

As with the question of peak efficiency for Penrose collisions, an
important factor in the observability of DM annihilation around black
holes is the question of the escape fraction for the resulting
reaction products. We showed above in Section \ref{sec:super} the planar
escape distribution for a selection of specially chosen
collisions. More general calculations of the escape probability have
been carried out in
\cite{Williams:2011,Banados:2011,Ogasawara:2017}. In short, the escape
fraction decreases as the distance from the black hole decreases (and
thus the center-of-mass energy increases). This is true for particles
plunging in from infinity. But for particles on stable, bound orbits,
the escape fraction can actually be quite large, on the order of
$90\%$ or more \cite{Schnittman:2015,Schnittman:2018}.

In most previous work on the subject, and in our own discussion above,
the DM population is generally divided into bound, and
unbound. However, when including self-interactions
(e.g., \cite{Spergel:2000}), these two populations can mix, giving
rise to new phenomenology and potentially greater enhancements of the
annihilation signal
\cite{Fields:2014,Shapiro:2014,Shapiro:2016}. Exotic DM particle
models with energy-dependent annihilation cross sections (e.g.,
\cite{Feng:2010,Chen:2013,Zurek:2014}) 
promise to make this field one of active research in the years to
come. 

Aside from DM annihilation, astrophysical applications of the
classical, collisional Penrose process are limited. In particular,
from everything we have seen in the review, extremely fine tuning and
multiple collisions would be required to get anywhere close to the
high-energy gamma-rays (or even cosmic rays) seen from many active
galactic nuclei. On the other hand, the high-energy emission that is
observed is likely indirectly related to the Penrose process, by
general coupling matter to the spin of the black hole. This can be
done far more efficiently
when employing large-scale fields, either in the form of
super-radiance \cite{Press:1972,Brito:2015}, or coupling the particles
directly to electromagnetic fields that in turn penetrate the black
hole horizon \cite{Blandford:1977,Wagh:1985}. Unfortunately, the high
efficiency of these mechanisms at creating gamma-rays only
serves to confuse and complicate any prospects of direct detection of
DM annihilation around otherwise quiescent galactic nuclei. We look
forward to the next generation of high-energy observatories that
will be able to circumvent these confusion sources with greater
sensitivity, and improved spatial and energy resolution. 

\begin{acknowledgements}
We thank Alessandra Buonanno, Francesc Ferrer, Ted
Jacobson, Henric Krawczynski, Tzvi Piran, Laleh Sadeghian, and Joe
Silk for helpful comments and discussion. A special thanks to the
editor of this Topical Collection, Emanuele Berti, for his
encouragement and patience.
\end{acknowledgements}

\bibliographystyle{spphys}       
\bibliography{penrose}   

\end{document}